\documentclass[reprint,aps,superscriptaddress,amsmath,longbibliography]{revtex4-2}
\usepackage{graphicx,txfonts,dcolumn,color}
\usepackage[colorlinks,linkcolor=blue,citecolor=blue,anchorcolor=blue,urlcolor=blue]{hyperref}

\begin{document}

\title{Iterative Site Percolation on Triangular Lattice}

\author{Ming Li}
\email{lim@hfut.edu.cn}
\affiliation{School of Physics, Hefei University of Technology, Hefei, Anhui 230009, China}

\author{Youjin Deng}
\email{yjdeng@ustc.edu.cn}
\affiliation{Hefei National Research Center for Physical Sciences at the Microscale, University of Science and Technology of China, Hefei 230026, China}
\affiliation{Department of Modern Physics, University of Science and Technology of China, Hefei, Anhui 230026, China}
\affiliation{Hefei National Laboratory, University of Science and Technology of China, Hefei, Anhui 230088, China}

\date{\today}

\begin{abstract}
The site percolation on the triangular lattice stands out as one of the few exactly solved statistical systems. By initially configuring critical percolation clusters of this model and randomly reassigning the color of each percolation cluster, we obtain coarse-grained configurations by merging adjacent clusters that share the same color. It is shown that the process can be infinitely iterated in the infinite-lattice limit, leading to an iterative site percolation model. We conjecture from the self-matching argument that percolation clusters remain fractal for any finite generation, which can even take any positive real number by a generalized process. Extensive simulations are performed, and, from the generation-dependent fractal dimension, a continuous family of previously unknown universalities is revealed.
\end{abstract}

\maketitle

\section{Introduction}

Percolation is a model of cluster formation that can show a phase transition~\cite{Stauffer1991}. The standard percolation model is defined on a lattice by randomly occupying sites or bonds with some probabilities. At the percolation threshold, fractal geometries can be constructed by the occupied sites or bonds. For a few lattices, particularly in two dimensions ($2$D), exact results can be obtained based on lattice duality/matching~\cite{Baxter1989,Sykes1964,Ziff2006}, Yang-Baxter integrability~\cite{Lieb1967,Baxter1972}, and local conformal invariance~\cite{Belavin1984,Friedan1984}. The fractal dimension $d_f=91/48$ of the percolation cluster in $2$D is predicted by Coulomb-gas arguments~\cite{Nienhuis1987,Saleur1987}, conformal field theory~\cite{Cardy1987,Francesco2012}, and stochastic Loewner evolution (SLE) theory~\cite{Kager2004,Cardy2005}, and is rigorously proved on the triangular lattice~\cite{Smirnov2001}. Thus, the universal properties of the percolation problem in $2$D are well understood, and the theoretical construction is propelled~\cite{Lawler2008,Jacobsen2009}.

Insights from percolation theory also enhance our understanding of cluster formation in numerous other systems~\cite{Newman2002,Li2014,Fan2018,Xie2021}. Across various situations where the percolation model is applied, it is commonly assumed that the organization of the medium under percolation rules occurs as a one-time event. That is, the percolation configuration is constructed by applying the percolation rule only once. In contrast, the actual organizational process of the medium is often iterative, with each new iteration aimed at reorganizing the clusters formed in the previous process rather than the original medium itself~\cite{Karrer2011,Hu2011,Granell2014,Massaro2014,Liu2015,Hu2016,Li2020,Hu2020}. In such an iterative process, an intriguing question arises: how does the system evolve in a new iteration when the previous iteration has already brought the system to a critical state?

The evolution of critical clusters in a new iteration can occur in two ways: dismantling them into smaller ones or merging them into larger ones. Dismantling clusters can be achieved by defining a percolation process independently for each cluster. Insights from percolation on critical but dense clusters~\cite{Liu2015} and on fractals~\cite{Havlin1983} suggest that a standard percolation transition with new universality classes is likely to be observed when a new iteration is taken on critical clusters. For merging critical clusters, a special case is to reorganize them in a scale-invariant manner, such as through the coarse-grained configuration in the celebrated renormalization group theory, for which the criticality maintains and the universality class remains unchanged. On the other hand, when critical clusters are merged by randomly placing additional occupied bonds, it is naturally expected that a giant cluster is formed to occupy a finite fraction of the whole lattice -- i.e., the long-range order emerges. This is because criticality corresponds to an unstable fixed point in the renormalization group flow~\cite{Ma2018}, where even minor disturbances can cause the system to deviate from this fixed point. However, there is no evidence to suggest that the merging of critical clusters must inevitably lead to a trivial state, with long-ranged order, if the merging rule is properly chosen.

The objective of this study is to formulate a simple model, termed \emph{iterative percolation}, in which critical percolation clusters are iteratively reorganized, and to investigate its percolation properties. We focus on the iterative site percolation on the triangular lattice, which begins with a critical site percolation configuration. For convenience, we call the occupied (unoccupied) sites and clusters black (white), and percolation clusters are constructed by adjacent sites with the same color, thus, any pair of adjacent clusters must have different colors. Then, by randomly reassigning the color of each cluster, the original adjacent clusters could share the same color, and, as a consequence, larger clusters are formed. This cluster-merging rule can be applied iteratively to the newly formed clusters, allowing for the sequential construction of new generations of percolation configurations.

From extensive simulation, we demonstrate that for any finite generations, such random merging of critical adjacent clusters does not yield a giant cluster, i.e., the percolating state. Instead, the system consistently maintains criticality, and the newly formed clusters have fractal structures, with generation-dependent fractal dimensions throughout the iterative process. Critical behaviors of various quantities, including susceptibility and cluster number density, follow the standard finite-size scaling theory for each generation. Notably, these critical behaviors do not belong to any previously identified universality classes. In the infinite-lattice limit, the iteration can continue infinitely, even extending to non-integer generations, thereby unveiling a continuous family of previously unknown universalities. According to the renormalization group theory, these observations imply a fascinating scenario in parameter space, where each point along the trajectory of successive iterations represents a fixed point, each possessing its own unique universality.

The remainder of this paper is organized as follows. In Sec.~\ref{sec-mo}, the model, observables, and finite-size scaling ansatz are introduced. Then, in Sec.~\ref{sec-sr}, the numeric results to show the generation-dependent universality are presented. A generalized model for any positive real generations is defined in Sec.~\ref{sec-gn}. We summarize the whole paper in Sec.~\ref{sec-con}.

\section{Model, observables, and finite-size scaling ansatz}   \label{sec-mo}

\subsection{Model}

We consider iterative percolation defined on a triangular lattice of linear size $L$ with periodic boundary conditions. Initially, at generation $n = 0$, each lattice site is sequentially visited and randomly colored white or black with a probability $p=1/2$. Any nearest-neighbor sites of the same color are then connected to form percolation clusters. Due to the self-matching property of triangular lattices~\cite{Sykes1964}, if there is no vertically wrapping cluster, there must be a horizontally wrapping cluster of the opposite color, and vice versa. Consequently, generation $n=0$ is precisely at the site-percolation threshold, and both white and black clusters are critical with the fractal dimension $91/48$.

\begin{figure}
\centering
\includegraphics[width=1.0\columnwidth]{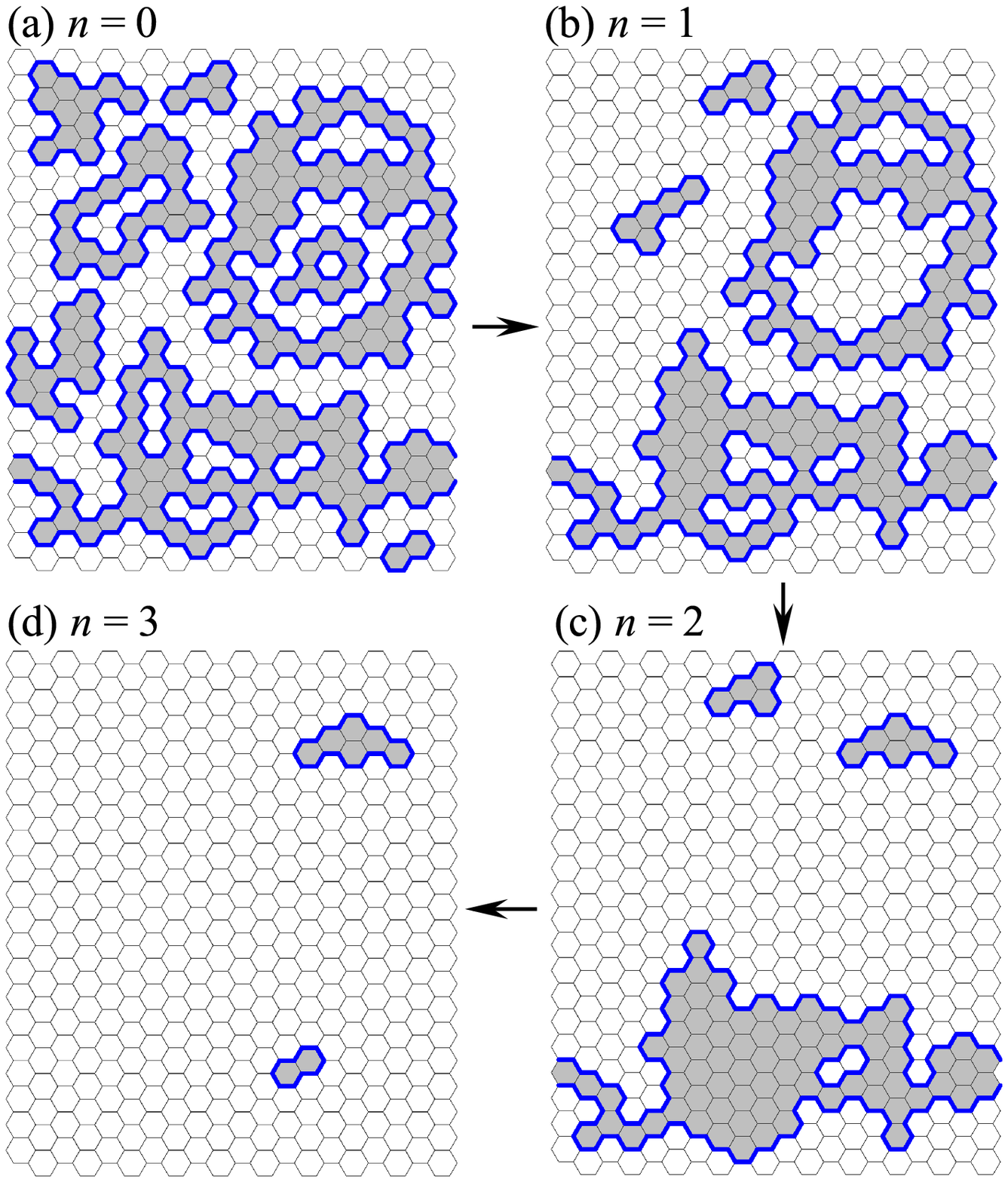}
\caption{(Color online) A schematic illustration, for generations from $n=0$ to $3$, of iterative site percolation on a triangular lattice of linear size $L=20$ with periodic boundary conditions. The loop representation on hexagonal lattices is utilized, where each elemental hexagonal face corresponds to a site in the triangular lattice. Within this loop representation, a cluster formed by adjacent elemental hexagonal faces of the same color should be entirely enclosed by another cluster of a different color, with the boundary represented by a loop highlighted in blue. The initial configuration (a) is generated by randomly coloring each elemental hexagonal face with either white or black, constituting a critical configuration. Subsequent generations $n=1$, $2$, and $3$ are depicted in (b)-(d), respectively. In each iteration, sites within the same cluster are reassigned a new color simultaneously, randomly chosen from white and black. After each iteration, any two adjacent clusters independently become the same color with a probability of $1/2$. Consequently, in a new generation, each loop is removed from the system with a probability of $1/2$. Note that, the iteration cannot create new loops, and the structures of preserved loops remain unchanged, suggesting a consistent fractal property of loops with the hull exponent $d_h=7/4$.}
\label{f1}
\end{figure}

Then, independently for each cluster, white or black color is randomly chosen, and assigned to all the sites within the cluster, regardless of their previous colors. As a consequence, with probability $1/2$, any pair of adjacent clusters comes out with the same color and they are merged into a larger percolation cluster. In other words, a new configuration of coarse-grained white and black clusters is obtained, giving generation $n=1$. Similarly, any generation $n+1$ can be constructed from generation $n$. In a finite system, there exists a maximum number of generations, denoted as $n_{max}$, after which the entire lattice is covered by either white or black. However, for infinite system, this color-flipping rule will never break the symmetry between white and black.

In Fig.~\ref{f1}, we provide a sketch of iterative percolation for generations from $n=0$ to $3$. Due to the duality between the triangular lattice and the hexagonal lattice, the configuration of site percolation on the triangular lattice can be equivalently obtained by coloring each elemental hexagonal face of a hexagonal lattice with either white or black. As depicted in Fig.~\ref{f1} (a), each cluster formed by the adjacent hexagonal faces of the same color should be completely enclosed in another cluster of the opposite color, with the boundary represented by a loop. Hence, this configuration is also referred to as the \emph{loop representation} of the percolation configuration~\cite{Duplantier1989}. Moreover, the duality between the triangular lattice and the hexagonal lattice ensures a strict one-to-one correspondence between the bonds in the hexagonal lattice and those in the triangular lattice. The bonds on the loops of the hexagonal lattice separate the clusters of different colors, and correspond to the inter-cluster bonds in the triangular lattice, while other other bonds are intra-cluster ones. Because of the equal and random distribution of white and black sites in generation $n=0$, the two types of bonds initially take an equal share.

It is worth noting that at criticality, the loops in the loop representation also exhibit fractal properties, often referred to as external perimeters or hulls, with a hull exponent of $d_h=7/4$~\cite{Saleur1987}. As depicted in Figs.~\ref{f1} (b)-(d), with the iteration processes, no new loops are created, and the number of loops decreases as clusters merge. Exactly, after an iteration, only half of the loops can be preserved, as any two adjacent clusters will independently display the same color with a probability of $1/2$. Furthermore, the structure of all the preserved loops remains unchanged, so that, the hull exponent $d_h=7/4$ should remain constant throughout the iteration.

\subsection{Observables}

In each generation, we sample the cluster information without distinction of white and black, including the size of the $k$-th largest cluster $\mathcal{C}_{k}$, the number $\mathcal{N}_s$ of clusters with size $s$, and the total number $\mathcal{N}_{inter}$ of inter-cluster bonds. To measure the susceptibility-like quantity, we assign an Ising-like auxiliary variable $\sigma=\pm1$ to white and black sites, respectively. Then, the Fourier-transformed susceptibility of the wave vector $\mathbf{k}$ can be defined as~\cite{Wittmann2014}
\begin{equation}
\chi_\mathbf{k} \equiv\frac{1}{L^2} \left\langle \left(\sum_j e^{i\mathbf{k}\mathbf{r}_j}\sigma_j\right)^2\right\rangle,    \label{eq-chik}
\end{equation}
where the sum $\sum_j$ runs over all the sites, $\mathbf{r}_j$ is the coordinate of site $j$, and $\langle \cdot\rangle$ refers to the statistical average.

With samples of independent realizations, we compute the following observables:
\begin{itemize}
  \item The sizes of the largest and the second largest clusters $C_1$ and $C_2$, computed as $C_{k}\equiv\langle{\mathcal C}_{k}\rangle$ with $k=1,2$.
  \item The probability distribution $F(x)$ of the largest-cluster size ${\mathcal C}_{1}$ across different realizations.
  \item The susceptibility $\chi_0$, corresponding to the second moment of cluster-size distribution $\chi_0 = \left\langle \sum_k \mathcal{C}_k^2 \right\rangle/{L^2}$, where the sum $\sum_k$ runs all the clusters.
  \item The Fourier-transformed susceptibility $\chi_1$ for the case of the smallest wave vector $\mathbf{k}=(2\pi/L,0)$ as defined in Eq.~(\ref{eq-chik}).
  \item The cluster number density $P(s',L)=\langle\mathcal{N}_s(\Delta s)\rangle/L^2\Delta s$, where $\mathcal{N}_s(\Delta s)$ represents the total number of clusters with size in the interval $[s,s+\Delta s)$. Here, the bin size $\Delta s$ is chosen as $\Delta s=a^k$ for the $k$-th bin from $s=1$, where $a>1$ is an adjustable parameter controlling the total number of bins, and $a=1.1$ is used in this paper. With this power-law growth bin size, the bin center is chosen as $s'=\sqrt{s(s+\Delta s)}$.
  \item The fraction of inter-cluster bonds $\rho=\langle\mathcal{N}_{inter}\rangle/3L^2$, where $3L^2$ is the total number of bonds on the triangular lattice.
  \item The maximum generation $n_{max}$ with a single cluster occupying the whole finite lattice.
\end{itemize}
Here, the angular brackets $\langle \cdot\rangle$ denote averaging across realizations.

To explore the finite-size scaling, the systems of different linear size from $L=2^4$ to $2^{14}$ are simulated. The simulation results are obtained by the average of $10^8$ realizations for systems of $L\leq 2^7$; for $2^8\leq L \leq 2^{14}$, the results are obtained by the average of $10^7\sim 10^4$ realizations.

\subsection{Finite-size scaling ansatz}

If iterative percolation generates critical configurations, according to percolation theory, observables $C_1$, $C_2$, $\chi_0$ and $\chi_1$ are expected to asymptotically diverge following power-law scaling with linear size $L$. Specifically, the finite-size scalings can be expressed as
\begin{align}
C_1 &\sim C_2 \sim L^{d_f},  \label{eq-cn}  \\
\chi_0 &\sim \chi_1 \sim L^{2d_f-d}, \label{eq-chi}
\end{align}
where $d=2$ is the spatial dimension, and $d_f$ is the fractal dimension. Moreover, the cluster number density $P(s,L)$ follows a scaling form given by
\begin{equation}
P(s,L)=s^{-\tau}\tilde{P}(s/L^{d_f}),
\end{equation}
where $\tau$ represents the Fisher exponent, and the hyperscaling relation dictates $\tau=1+d/d_f$.

The system at generation $n=0$ represents a critical configuration of standard percolation, thus, all the aforementioned finite-size scalings hold true with $2$D universality, characterized by $d_f=91/48$ and $\tau=187/91$. For $n>0$, we will demonstrate that these finite-size scalings can still be observed, with generation-dependent critical exponents. To determine these exponents, we fit the Monte Carlo data for various linear sizes $L$ to the finite-size scaling ansatz given by
\begin{equation}
Q(L) = L^{Y_Q} (a_0+a_1L^{-\omega_1}+a_2L^{-\omega_2}),  \label{eq-fit}
\end{equation}
where the exponent $Y_Q$ describes the leading finite-size behavior, and $\omega_i$ with $i=1$ and $2$ account for the finite-size corrections. During the fitting process, a lower cutoff $L\geq L_{\text{min}}$ is imposed on the data points, and we systematically assess the goodness and stability of fits by increasing $L_{\text{min}}$. Generally, our preferred fit result corresponds to the smallest $L_{\text{min}}$ for which a chi-square per degree of freedom around $1$ is obtained. Moreover, increasing $L_{\text{min}}$ does not cause a significant increase in the chi-square per degree of freedom.

\section{Simulation results}   \label{sec-sr}

\subsection{Logarithmic divergence of maximum generation}

\begin{figure}
\centering
\includegraphics[width=1.0\columnwidth]{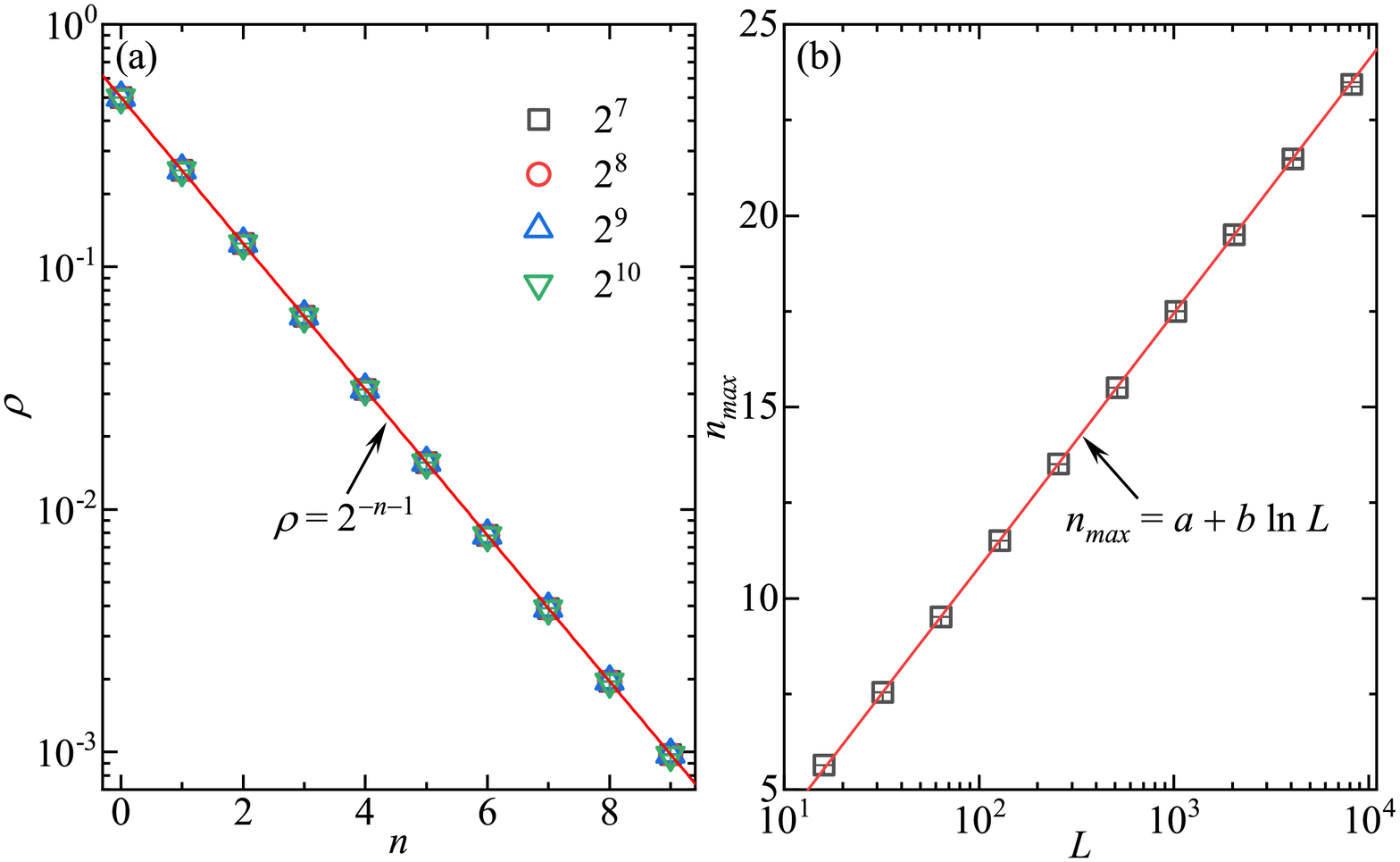}
\caption{(Color online) Exactly solvable quantities in iterative site percolation on triangular lattices. (a) The density of inter-cluster bonds $\rho$ as a function of generation $n$ for different linear sizes $L$. The $\rho$ data for different $L$ are precisely on top of each other, due to the absence of $L$-dependence. The line represents the exact solution $\rho(n)=2^{-n-1}$. (b) The logarithmic divergence of the maximum generation $n_{max}$ that all the sites of a finite system belong to a single cluster. The line shows the function $a+b\ln L$ with the fit results $a=-2.45(3)$ and $b=2.87(1)$, where $b$ agrees well with the exact value $b=2/\ln 2\approx2.885$.} \label{f2}
\end{figure}

As the definition of iterative percolation, in each iteration, any two adjacent clusters merge independently with a probability of $1/2$. In the loop representation, this implies that each loop will be removed with a probability of $1/2$ in a new generation, irrespective of the percolation configuration. Moreover, a one-to-one correspondence exists between the bonds of loop element on the hexagonal lattice and the inter-cluster bonds on the triangular lattice, thereby, the number of inter-cluster bonds decreases by half for each iteration. Consequently, we find an exact expression for the density of inter-cluster bonds for any generation $n$,
\begin{equation}
\rho(n)=2^{-n-1},  \label{eq-rho}
\end{equation}
where the initial density of inter-cluster bonds $\rho(0)=1/2$ is used. It is worth noting that the above analysis is devoid of any approximations and is also independent of the system size. Therefore, for a given generation $n$, the data of $\rho(n)$ from different system sizes exhibit perfect overlap, as shown in Fig.~\ref{f2} (a).

For finite systems, the iterative process would eventually end at a maximum generation $n_{max}$ with a single cluster occupying the whole lattice. Approaching $n_{max}$, the total number of inter-cluster bonds should become finite, given by
\begin{equation}
3L^2 \rho(n_{max}) \sim \mathcal{O}(1).    \label{eq-nmax}
\end{equation}
Here, $3L^2$ is the total number of bonds on the triangular lattice. Substituting Eq.~(\ref{eq-rho}) into Eq.~(\ref{eq-nmax}), the maximum generation $n_{max}$ can be estimated as
\begin{equation}
n_{max}=\frac{2}{\ln 2}\ln L + \mathcal{O}(1).
\end{equation}
This agrees well with the simulation result in Fig.~\ref{f2} (b), and shows that, in the infinite-lattice limit, iterative percolation can be well defined for any finite generation.

\subsection{Fractal dimension}

\begin{figure}
\centering
\includegraphics[width=1.0\columnwidth]{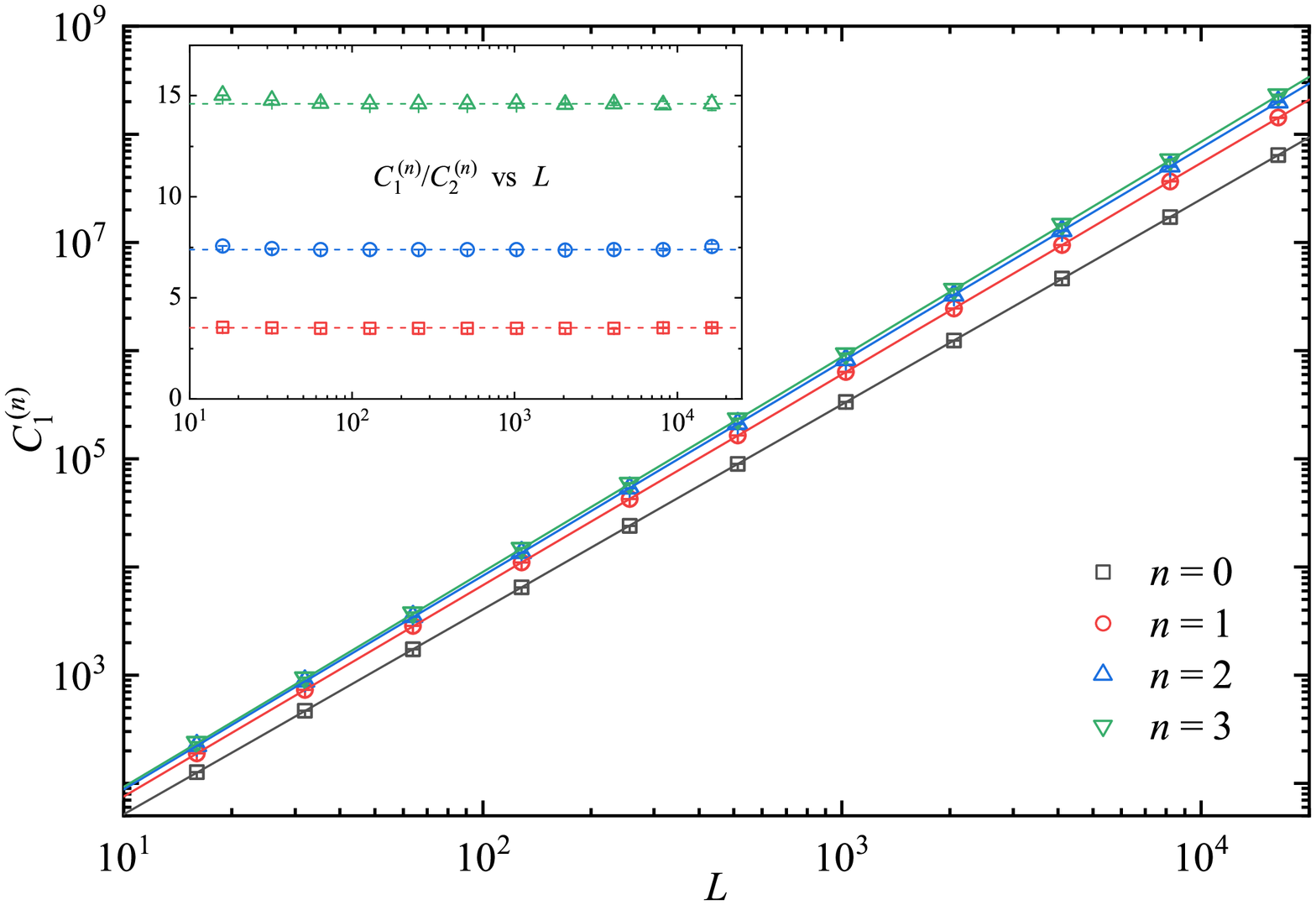}
\caption{(Color online) The size of the largest cluster $C_1^{(n)}$ as a function of linear size $L$ for various generations $n$. The lines represent the fit results of fractal dimensions in Table~\ref{t1}. The inset displays the ratio $C_1^{(n)}/C_2^{(n)}$ as $L$ increases, indicating a size-independent value (dashed lines) for each generation. This observation suggests that the two largest clusters have the same fractal dimension.}       \label{f3}
\end{figure}

\begin{table}[b]
\caption{The fit results of the fractal dimensions $d_f^{\,(n)}$ for generations from $n=0$ to $5$, as obtained from the sizes of the largest and the second-largest clusters $C_1\sim C_2\sim L^{d_f}$, and, the susceptibility and the Fourier-transformed susceptibility $\chi_0\sim\chi_1\sim L^{2d_f-d}$. The results are obtained by fitting the data to the scaling ansatz Eq.~(\ref{eq-fit}), with $Y_Q=d_f$ for $C_1$ and $C_2$, and with $Y_Q=2d_f-2$ for $\chi_0$ and $\chi_1$. In the fit, only one correction term ($a_2=0$) is considered with free $a_1$ and $\omega_1$. For each iteration, the estimated values of $d_f$ from different quantities are consistent with each other, suggesting that the criticality is well-defined. For different iterations, the values of $d_f$ vary significantly, exceeding the error bars by a considerable margin, suggesting generation-dependent universalities.}   \label{t1}
\begin{ruledtabular}
\begin{tabular}{ccccccc}
$n$    & $0$  & $1$  & $2$  & $3$  & $4$  & $5$ \\
\hline
%$C_1$     & $1.89584(5)$    & $1.95126(4)$    & $1.97638(5)$    & $1.98836(4)$     & $1.99420(5)$     & $1.99712(2)$    \\
$C_1$     & $1.89584(5)$    & $1.9513(1)$   & $1.9764(1)$   & $1.9884(1)$    & $1.9942(1)$    & $1.9971(1)$    \\
$C_2$     & $1.89581(6)$    & $1.9514(2)$   & $1.9762(3)$   & $1.9882(3)$    & $1.9940(4)$    & $1.9970(6)$   \\
$\chi_0$  & $1.8954(3)~~$   & $1.9514(2)$   & $1.9765(2)$   & $1.9884(2)$    & $1.9942(1)$    & $1.9971(2)$    \\
$\chi_1$  & $1.8958(2)~~$   & $1.9513(3)$   & $1.9762(2)$   & $1.9884(7)$    & $1.9941(5)$    & $1.9967(7)$
\end{tabular}
\end{ruledtabular}
\end{table}

The observation in Fig.~\ref{f1} reveals that while the number of loops decreases with subsequent generations, their fractal structures, i.e., hull exponent, remains unchanged. This implies that despite the reduction in loop numbers, successive generations remain at criticality. Moreover, the self-matching property of triangular lattices also suggests that the system is always at criticality, as long as the symmetry between black and white is not broken. Note that, despite of the existence of effective ``interactions" between neighboring sites for $n>0$, the self-matching argument still holds true. To further verify whether the newly generated configuration is at criticality, we can examine the finite-size scaling of the largest and second-largest clusters $C_1$ and $C_2$. From Fig.~\ref{f3}, it is evident that the finite-size scaling of $C_1^{(n)}$ for generation $n\geq1$ can still be well described by the finite-size scaling of Eq.~(\ref{eq-cn}). This indicates that the configuration of generation $n\geq1$ is still at criticality.

By fitting the data of $C_1^{(n)}$ to finite-size scaling ansatz Eq.~(\ref{eq-fit}) with only one correction term ($a_2=0$), and both $a_1$ and $\omega_1$ are free, steady fit results can be obtained, as listed in the first line of Table~\ref{t1} for generations from $n=0$ to $5$. It is noteworthy that even without including any correction terms in Eq.~(\ref{eq-fit}), i.e., $a_1=0$ and $a_2=0$, consistent fit results can be achieved by using a larger cutoff $L_{\text{min}}$.

At criticality, not only the largest cluster but also other clusters, such as the second largest cluster, should be fractal with the same fractal dimension as $C_1^{(n)}$. The fractal dimensions of the second largest clusters, determined by fitting the data of $C_2^{(n)}$ to the finite-size scaling ansatz Eq.~(\ref{eq-fit}) with $a_2=0$, are also presented in Table~\ref{t1} for generations from $n=0$ to $5$. As expected, these fractal dimensions are well consistent with those obtained from the fit of $C_1^{(n)}$ within the error bars, which is further confirmed by the size-independent values of the ratios $C_1^{(n)}/C_2^{(n)}$ shown in the inset of Fig.~\ref{f3}. These observations reinforce the well-defined criticality for each generation.

\begin{figure}
\centering
\includegraphics[width=1.0\columnwidth]{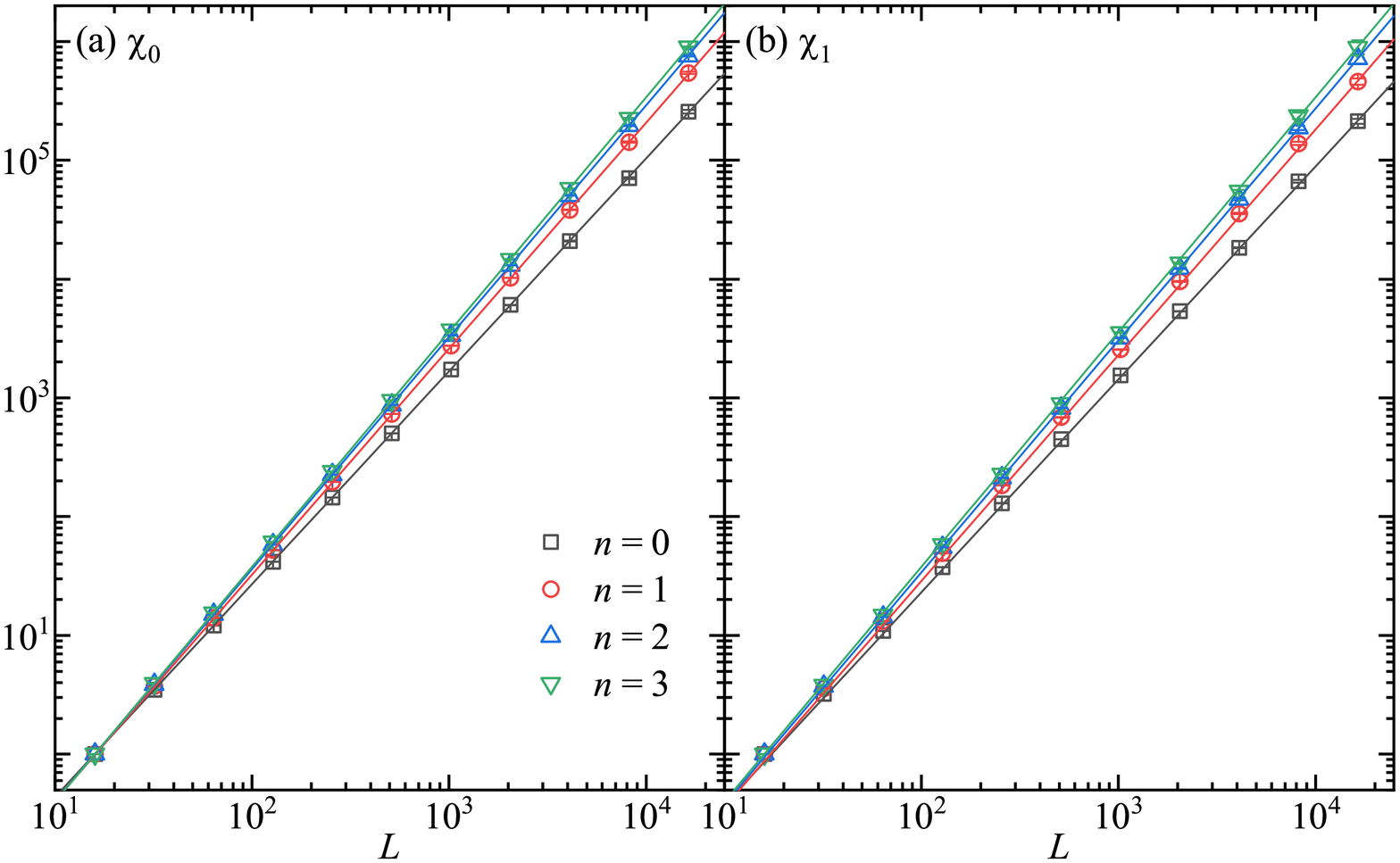}
\caption{(Color online) The susceptibility $\chi_0^{(n)}$ (a) and the Fourier-transformed susceptibility $\chi_1^{(n)}$ (b) as a function of linear size $L$ for different generations $n$. The lines show the scaling $\chi^{(n)}\sim L^{2d_f^{\,(n)}-2}$, with the fit results of $d_f^{\,(n)}$ given in Table~\ref{t1}. To visually illustrate the distinct finite-size scalings for each generation, all the data points are collectively rescaled such that the first data points of different generations $n$ have the same value of $1$.} \label{f4}
\end{figure}

Further, as the finite-size scaling of Eq.~(\ref{eq-chi}), the fractal dimension of any generation can be also obtained by fitting the data of the susceptibility $\chi_0$ and the Fourier-transformed susceptibility $\chi_1$ to the scaling ansatz Eq.~(\ref{eq-fit}) with $Y_Q=2d_f-2$. With only one correction term ($a_2=0$) and free $a_1$ and $\omega_1$, steady fit results can be found as listed in Table~\ref{t1}. According to the identical fractal dimension from the data of $C_1^{(n)}$, $C_2^{(n)}$, $\chi_0^{(n)}$ and $\chi_1^{(n)}$ in Table~\ref{t1}, the well-defined criticality is immediately verified for each generation. Moreover, by examining the finite-size scaling of the ratios $L^2/C_1^{(n)}$ and $C_1^{(n)}/C_1^{(n-1)}$, we can find more numerical evidences for the criticality of iterative percolation, see Appendix \ref{app-ne} for details.

\subsection{Probability distribution of the largest-cluster size}

\begin{figure}
\centering
\includegraphics[width=1.0\columnwidth]{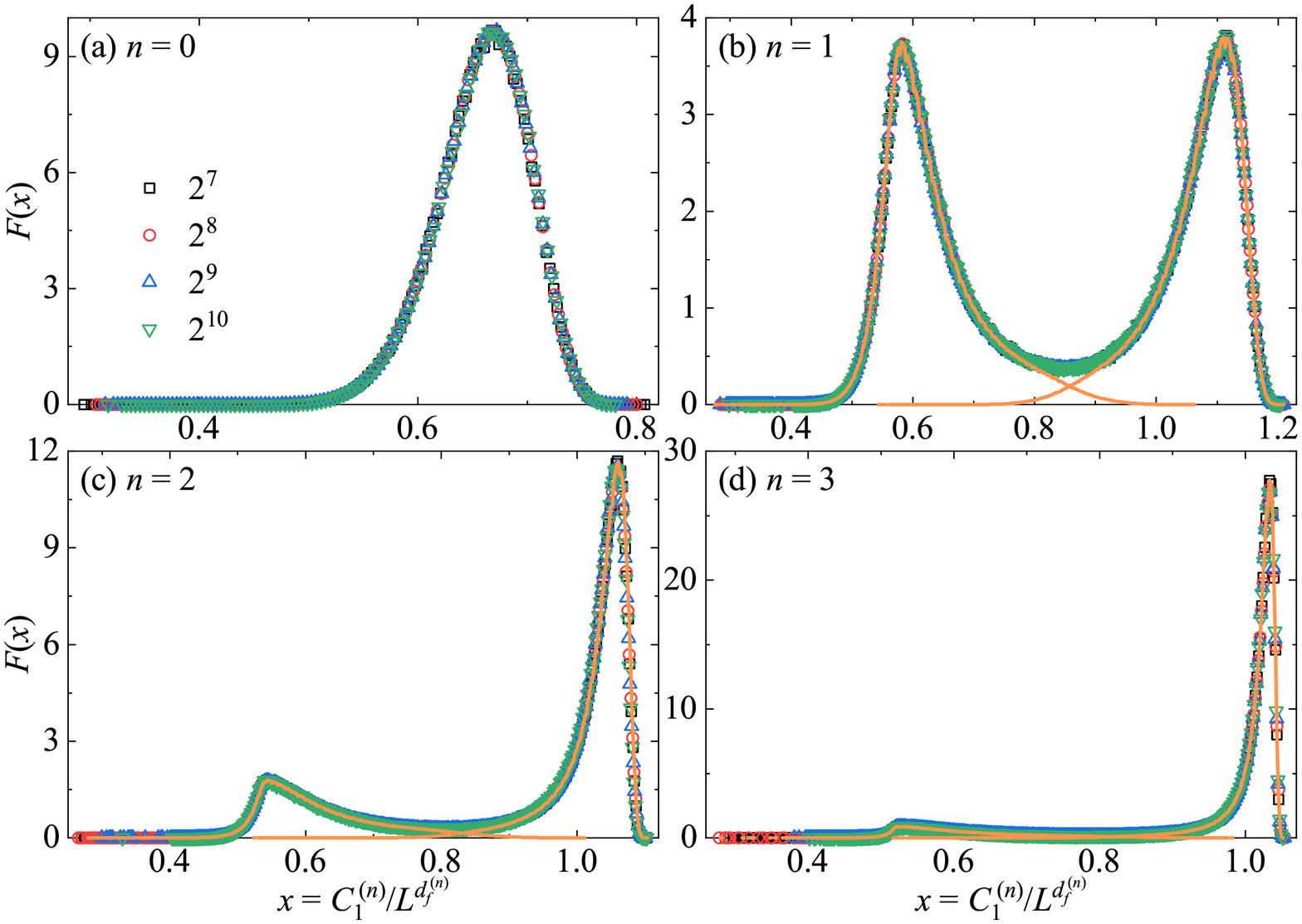}
\caption{(Color online) The probability density $F(x)$ of the size of the largest cluster for different linear sizes $L$ and different generations $n$. For $n>0$, the probability density $F(x)$ shows a bimodal distribution, however, a good data collapse for the data of different $L$ is obtained for the whole range of $x$ by defining $x\equiv C_1^{(n)}/L^{d_f^{\,(n)}}$. By whether the two largest clusters of generation $n=0$ are contained within a cluster of the current generation, the bimodal distribution $F(x)$ is perfectly split into two unimodal distributions, see the orange curves for an example of $L=2^7$.} \label{f5}
\end{figure}

The results in Table~\ref{t1} also suggest that the fractal dimensions of successive generations are distinct, as the differences are much larger than the error bars (see Appendix \ref{app-ne} for the well-defined finite-size scaling of $C_1^{(n)}/C_1^{(n-1)}$). To further confirm the unique fractal dimension $d_f^{\,(n)}$ of each generation, the rescaled probability density of the size of the largest cluster $F(x)$ is displayed in Fig.~\ref{f5} for generations from $n=0$ to $3$. By defining $x\equiv C_1^{(n)}/L^{d_f^{\,(n)}}$ from the fit results $d_f^{\,(n)}$ in Table~\ref{t1}, a nice data collapse of $F(x)$ for different $L$ can be achieved. This suggests that, for any generation, the probability density of the size of the largest cluster can be described by a universal function $F(x)$, indicating the well-defined fractal dimension for each generation. Moreover, the form of the rescaled probability density $F(x)$ of iterative percolation varies with generations, and generally shows a bimodal distribution for $n>0$. This means that the universal function $F(x)$ is different from generation to generation, so that, the generation-dependent universality of iterative percolation is further confirmed.

Additionally, the bimodal distribution of $F(x)$ for $n>0$ in Fig.~\ref{f5} can be understood as follows. Roughly speaking, the right peak corresponds to the samples where the two largest clusters of generation $n=0$ have merged in the current generation, while the left peak represents other samples. When filtering the data of $\mathcal{C}_1$ based on this condition, the bimodal distribution $F(x)$ perfectly splits into two unimodal distributions, as depicted by the orange curves in Fig.~\ref{f5}. As the iteration progresses, more samples will fall into the case of the right peak. This clarifies the gradual diminishment of the left peak with increasing $n$. Importantly, this phenomenon does not affect the fractal geometry of the newly generated clusters, and data collapse can be achieved across the entire range of $x=C_1^{(n)}/L^{d_f^{\,(n)}}$.

\subsection{Cluster number density}

\begin{figure}
\centering
\includegraphics[width=1.0\columnwidth]{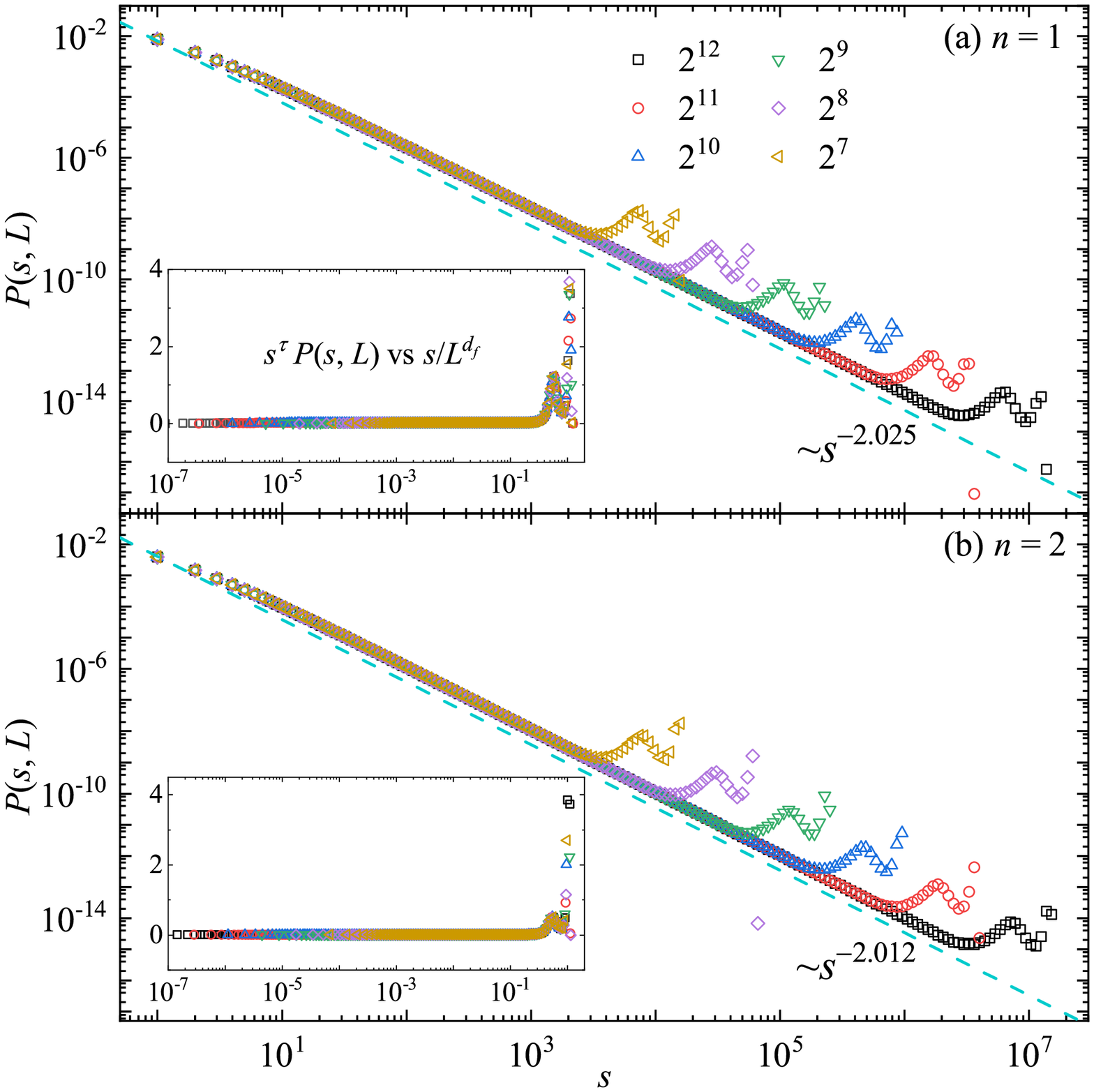}
\caption{(Color online) The cluster number densities $P(s,L)$ of $n=1$ (a) and $n=2$ (b) for different linear sizes $L$. The dashed lines show the Fisher exponent predicted by the hyperscaling relation $\tau^{(1)}=1+d/d_f^{\,(1)}\approx 2.025$ and $\tau^{(2)}=1+d/d_f^{\,(2)}\approx 2.012$ from the fit results of $d_f^{(n)}$ in Table~\ref{t1}, respectively. The inset shows $s^\tau P(s,L)$ versus $s/L^{d_f}$ for different $L$, where the data of different system sizes collapse on top of each other, and show a horizontal line for $s/L^{d_f}\ll 1$, suggesting a well-defined cluster number density for each generation.} \label{f6}
\end{figure}

The cluster number density $P(s,L)$ is a crucial quantity in percolation, also known as cluster-size distribution. At criticality, $P(s,L)$ follows a scaling form given by $P(s,L)=s^{-\tau}\tilde{P}(s/L^{d_f})$ with the hyperscaling relation $\tau=1+d/d_f$. In Fig.~\ref{f6}, we plot $P(s,L)$ of generations $n=1$ and $2$ versus cluster size $s$ for increasing $L$. As we can see, a power-law distribution of $P(s,L)$ is expected for $L\to\infty$, and the Fisher exponents $\tau^{\,(n)}$ are also generation-dependent, which is consistent with the value obtained by the hyperscaling relation $\tau^{\,(n)}=1+d/d_f^{\,(n)}$ from the fit result of $d_f^{\,(n)}$ in Table~\ref{t1}. Note that $\tilde{P}(x)$ is a universal function that is independent of the system size, thus by plotting $s^{\tau}\tilde{P}(x)$ as a function of $x=s/L^{d_f}$, the data of different system sizes should collapse on top of each other, and show a horizontal line for $x\ll1$. This behavior is confirmed by the insets of Fig.~\ref{f6}. All these suggest a well-defined cluster number density for each generation. Thus, iterative percolation is not a trick to simply evolve the fractal geometry of the largest cluster, instead, the system is actually at criticality, as all the clusters show a key feature of criticality -- scale invariance.

\begin{figure}
\centering
\includegraphics[width=1.0\columnwidth]{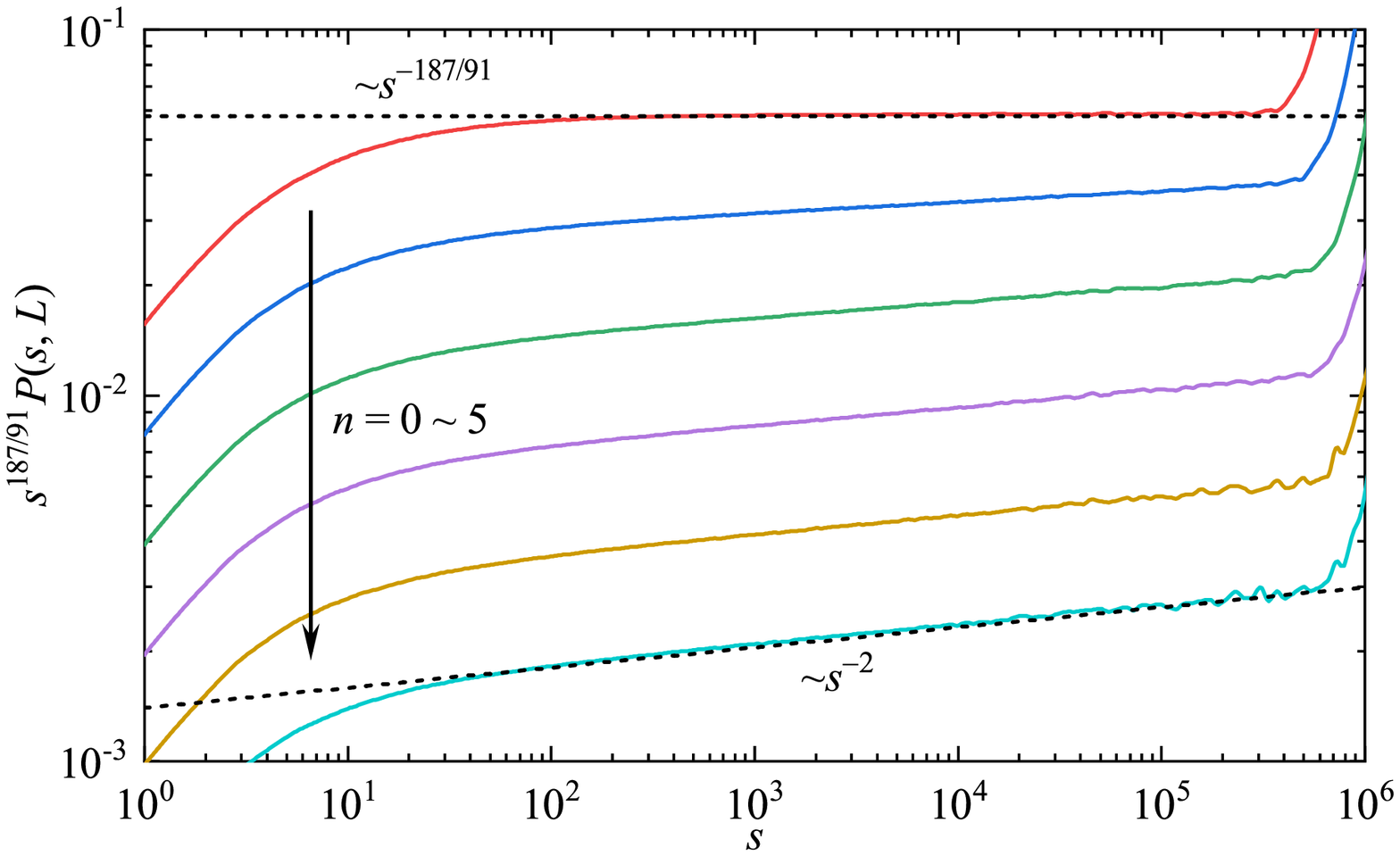}
\caption{(Color online) The cluster number density rescaled by $s^{187/91}$, namely, $s^{187/91}P(s,L)$, for generations from $n=0$ to $5$. Generation $n=0$ belongs to the $2$D percolation universality that $P(s,L)\sim s^{-187/91}$, thus, $s^{187/91}P(s,L)$ is parallel to the horizontal axis indicated by the dashed line. The deviations from the horizontal line for small and large $s$ are due to the finite system size. As generation increases, $s^{187/91}P(s,L)$ becomes significantly tilted with respect to a horizontal line, suggesting a different Fisher exponent, which approaches $2$ for $n\to\infty$ indicated by the dashed line. The triangular lattice used in the simulation has side length $L=2^{11}$.} \label{f7}
\end{figure}

According to the hyperscaling relation $\tau^{\,(n)}=1+d/d_f^{\,(n)}$, it is anticipated that the Fisher exponent is $\tau^{\,(\infty)}=2$, as $d_f^{\,(n)}\to d$ for $n\to\infty$. Notably, the Fisher exponent of $2$D percolation universality is $\tau^{\,(0)}=187/91\approx2.055$, a value close to the infinite-generation limit $\tau^{\,(\infty)}=2$. To elucidate the evolution of the Fisher exponent as the iteration progresses, we plot $s^{187/91}P(s,L)$ as a function of $s$ for generations from $n=0$ to $5$ in Fig.~\ref{f7}. We focus solely on the evolution of the Fisher exponent; therefore, only the middle region of $P(s,L)$ is displayed. For the $2$D percolation universality (generation $n=0$), the middle region of $s^{187/91}P(s,L)$ is evidently parallel to the horizontal axis. As the generation increases, the middle region of $s^{187/91}P(s,L)$ becomes noticeably inclined with respect to a horizontal line, and for $n=5$, it is nearly parallel to the line for $\tau=2$. Consequently, the evolution of the Fisher exponent from the $2$D percolation universality $187/91$ to $2$ illustrated in Fig.~\ref{f7} underscores the generation-dependent universality of iterative percolation.

\section{Generalization}  \label{sec-gn}

The cluster merging rule of iterative percolation can be equivalently described as flipping the color of each cluster with a probability of $1/2$. More generally, one can construct a new generation by flipping the color of each cluster with a probability $q\in(0,1)$. Due to the symmetry between white and black, flipping the color of each cluster with probability $q$ is equivalent to flipping the color of each cluster with probability $1-q$. Thus, we need only consider the case of $q\leq1/2$. Applying this rule with $q=1/2$ to generation $n$, a new configuration of integer generation $n+1$ will be generated as discussed above. For $q<1/2$, the result corresponds to a non-integer generation $n+\epsilon$, where $0<\epsilon<1$. The value of $\epsilon$ is determined by the probability $q$.

As discussed for Fig.~\ref{f2} (a), when the generation increases by one, each inter-cluster bond is preserved with a probability of $2^{-1}$. Naturally, this probability becomes $2^{-\epsilon}$, when the generation increases by a non-integer value $\epsilon$. In other words, two adjacent clusters stay separate with a probability of $2^{-\epsilon}$ for a generation increment $\epsilon$. On the other hand, for a color flipping probability $q$, two adjacent clusters maintain different colors with a probability of $q^2+(1-q)^2$, corresponding to the scenario where both or neither of the two adjacent clusters flip their colors. From this, we can establish a relation between $q$ and $\epsilon$, namely, $q^2+(1-q)^2=2^{-\epsilon}$. Hence, by flipping each cluster of generation $n$ with probability $q$, it results a non-integer generation $n+\epsilon$, where $\epsilon$ is given by
\begin{equation}
\epsilon=\frac{\ln\left[q^2+(1-q)^2\right]}{\ln \frac{1}{2}}.   \label{eq-eps}
\end{equation}
With this, any positive real generations can be defined. For $q=1/2$, Eq.~(\ref{eq-eps}) reduces to $\epsilon=1$, corresponding to the integer increment of generations.

\begin{figure}
\centering
\includegraphics[width=1.0\columnwidth]{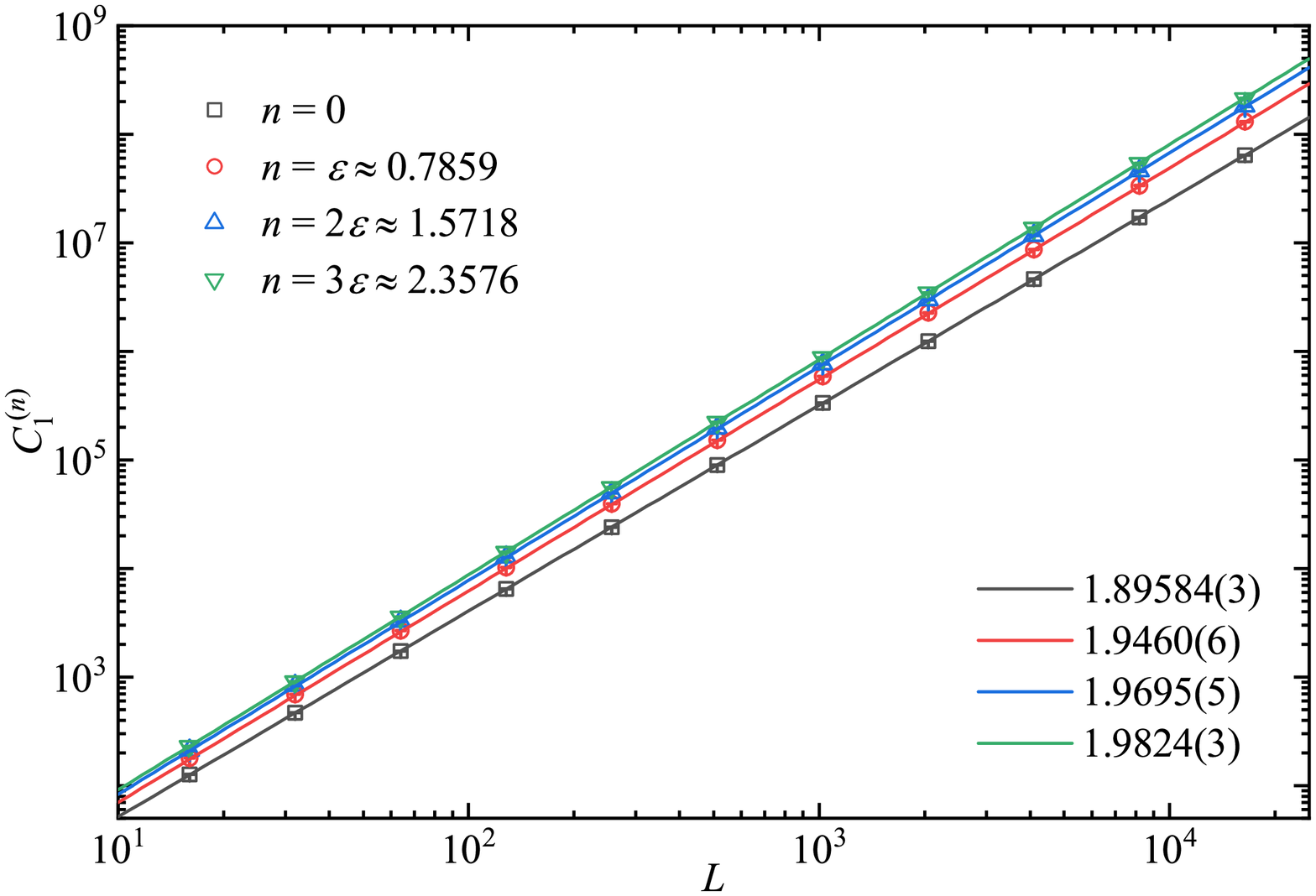}
\caption{(Color online) The size of the largest cluster $C_1$ as a function of linear size $L$ for some non-integer generations. These generations are generated by iteratively flipping the color of each cluster with a probability of $q=0.3$ starting from generation $n=0$, with each iteration increases the generation by approximately $\epsilon\approx0.7859$. The lines represent the fractal dimensions obtained by fitting the data to the scaling ansatz Eq.~(\ref{eq-fit}) with only one correction term ($a_2=0$), and free $a_1$ and $\omega_1$.} \label{f8}
\end{figure}

As an example, we iteratively flip the color of each cluster with probability $q=0.3$ starting from generation $n=0$. According to Eq.~(\ref{eq-eps}), each iteration increases the generation by approximately $\epsilon\approx0.7859$. The finite-size scalings of $C_1$ for some non-integer generations are shown in Fig.~\ref{f8}, where they are well described by the scaling of Eq.~(\ref{eq-cn}). Specifically, for generations $n=0.7859$, $1.5718$, and $2.3576$, the fit results give the fractal dimensions $d_f^{\,(n)}=1.9460(6)$, $1.9695(5)$, and $1.9824(3)$, respectively. It can be seen that the fractal dimension of a non-integer generation $n$ lies between those of generations $\lfloor n\rfloor$ and $\lceil n\rceil$ (see Table~\ref{t1} and Fig.~\ref{f8}), where $\lfloor n\rfloor$ and $\lceil n\rceil$ mean the integers by rounding down and up to $n$, respectively. This observation suggests that for continuously variable generations $n$, the universalities with continuously variable fractal dimensions from $91/48$ to $2$ are expected. In terms of renormalization group flow, this generalized model produces a special trajectory in parameter space. Each point along this trajectory represents a fixed point with its own unique universality, and the critical exponent changes continuously along this line.

\section{Conclusion}   \label{sec-con}

In this paper, we introduce an iterative site percolation model. Starting from the initial generation, set to a critical configuration of site percolation on triangular lattices, all subsequent generations maintain criticality with distinct fractal dimensions. Additionally, other typical critical phenomena, such as the power-law distribution of cluster sizes (with the Fisher exponent satisfying the standard hyper-scaling relationship), the divergence of susceptibility, and the scaled probability distribution of the largest cluster, can all be observed with generation-dependent critical exponents. Using the loop representation on hexagonal lattices, we derive the exact solution for the density of inter-cluster bonds, elucidating the logarithmic divergence of the maximum generation. This indicates that iterative site percolation can be well-defined for any finite number of generations and further generalized to any positive real generation, revealing a continuous family of previously unknown universalities.

The criticality of iterative percolation can be well demonstrated by introducing a bond occupied probability $p_b$ within each cluster of iterative percolation (see Appendix \ref{app-bsp} for details). When $p_b<1$, it makes the system deviate from criticality, which allows us to measure the correlation-length exponent $\nu$ for each generation. The numeric results suggest that $\nu$ for such a bond-site percolation is generation-independent, and always takes the value $\nu=4/3$ of $2$D percolation universality. Note that, the hull exponent $d_h=7/4$ also remains unchanged with iterations, as demonstrated in Fig.~\ref{f1}. These indicate that although different generations have their own universalities, they still exhibit some essential characteristics of $2$D percolation.

This iterative percolation process can be broadly defined by iteratively coloring the critical clusters of various models. A natural extension is to apply a similar cluster coloring rule to critical clusters in bond percolation and in the random-cluster representation of the Potts model, across various lattice structures. When considering only a single iteration, this concept is linked to the so-called fuzzy Potts model~\cite{Maes1995}, also known as the divide and color model~\cite{Haeggstroem2001}, which has garnered significant attention from probability theorists.

Our findings unveil an intriguing physical depiction: in the renormalization group flow, the system consistently resides atop fixed points throughout the iterative process. This poses an open challenge for developing a theoretical understanding of these universality classes with continuously variable critical exponents, particularly within the framework of SLE theory, which has seen rapid advancement in recent years~\cite{Smirnov2001,Kager2004,Cardy2005,Nolin2023}. Therefore, delving into the physical picture of iterative percolation could hold fundamental significance in statistical physics, condensed matter physics, and quantum field theory.

\section*{Acknowledgements}

The authors acknowledge helpful discussions with Jingfang Fan and Hao Hu. The research was supported by the National Natural Science Foundation of China under Grant No.~12275263, the Innovation Program for Quantum Science and Technology under Grant No.~2021ZD0301900, and Natural Science Foundation of Fujian province of China under Grant No.~2023J02032. The research of M.L. was also supported by the Fundamental Research Funds for the Central Universities (No.~JZ2023HGTB0220).

\appendix

\section{Numerical evidences for ruling out other seemingly plausible scenarios}  \label{app-ne}

\begin{figure}
\centering
\includegraphics[width=1.0\columnwidth]{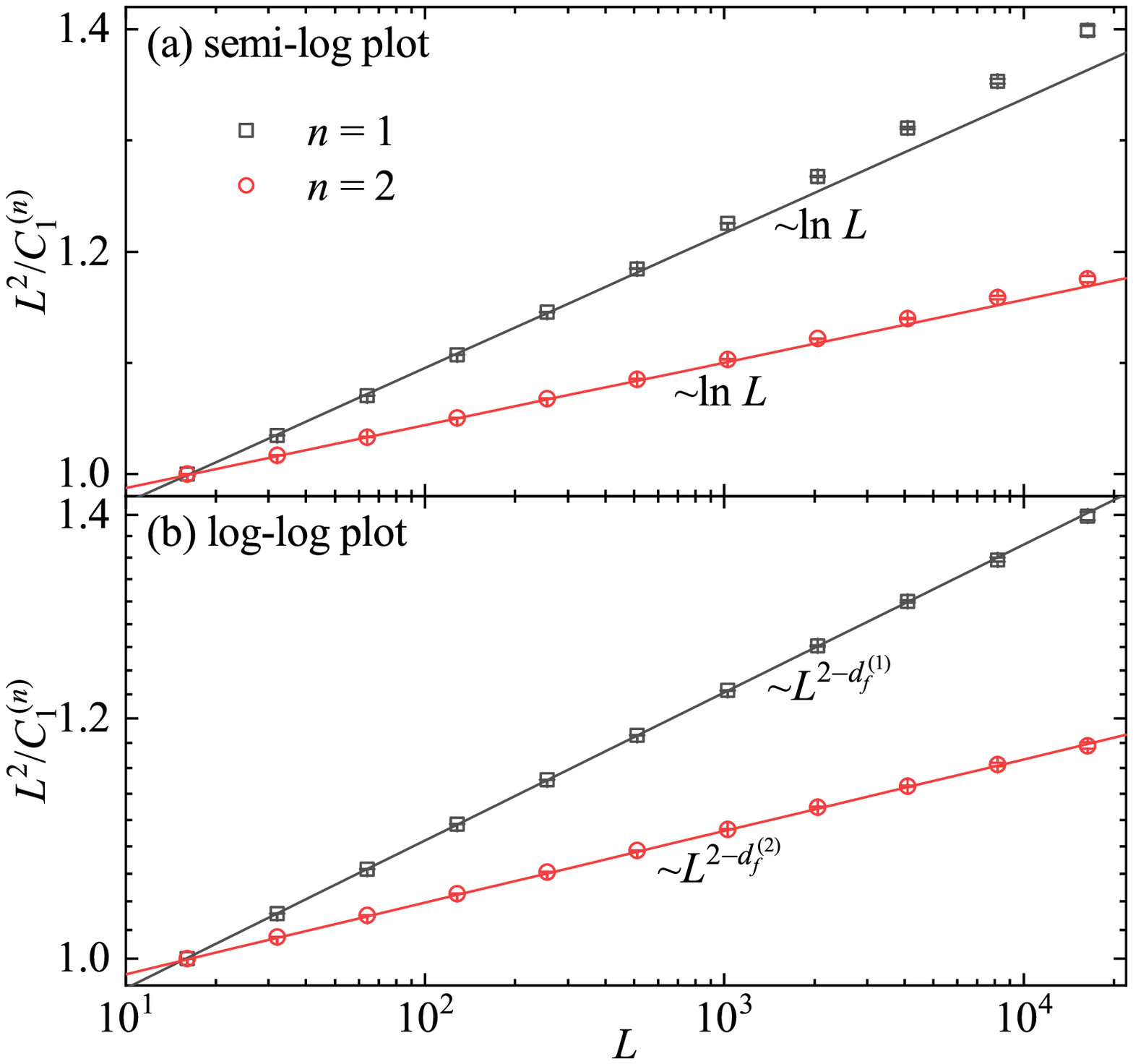}
\caption{(Color online) The finite-size scaling to demonstrate that the growth of $L^2/C_1^{(n)}$ for increasing system size is well described by a power-law function (b) rather than a logarithmic function (a), for generations $n=1$ and $2$. (a) In the semi-log plot, the ratio $L^2/C_1^{(n)}$ deviates upward from a straight line ($\sim\ln L$) for both $n=1$ and $2$, suggesting that the finite-size behaviors of $L^2/C_1^{(n)}$ cannot be described by logarithmic growth. (b) In the log-log plot, the same data as in (a) exhibit nice power-law behaviors indicated by the straight lines, representing the scaling $\sim L^{2-d_f^{\,(n)}}$ with the fractal dimensions from Table~\ref{t1}. To demonstrate generations $n=1$ and $2$ simultaneously, all the data are rescaled collectively to ensure that the first data point of different $n$ have the same value of $1$.} \label{f9}
\end{figure}

\begin{figure}
\centering
\includegraphics[width=1.0\columnwidth]{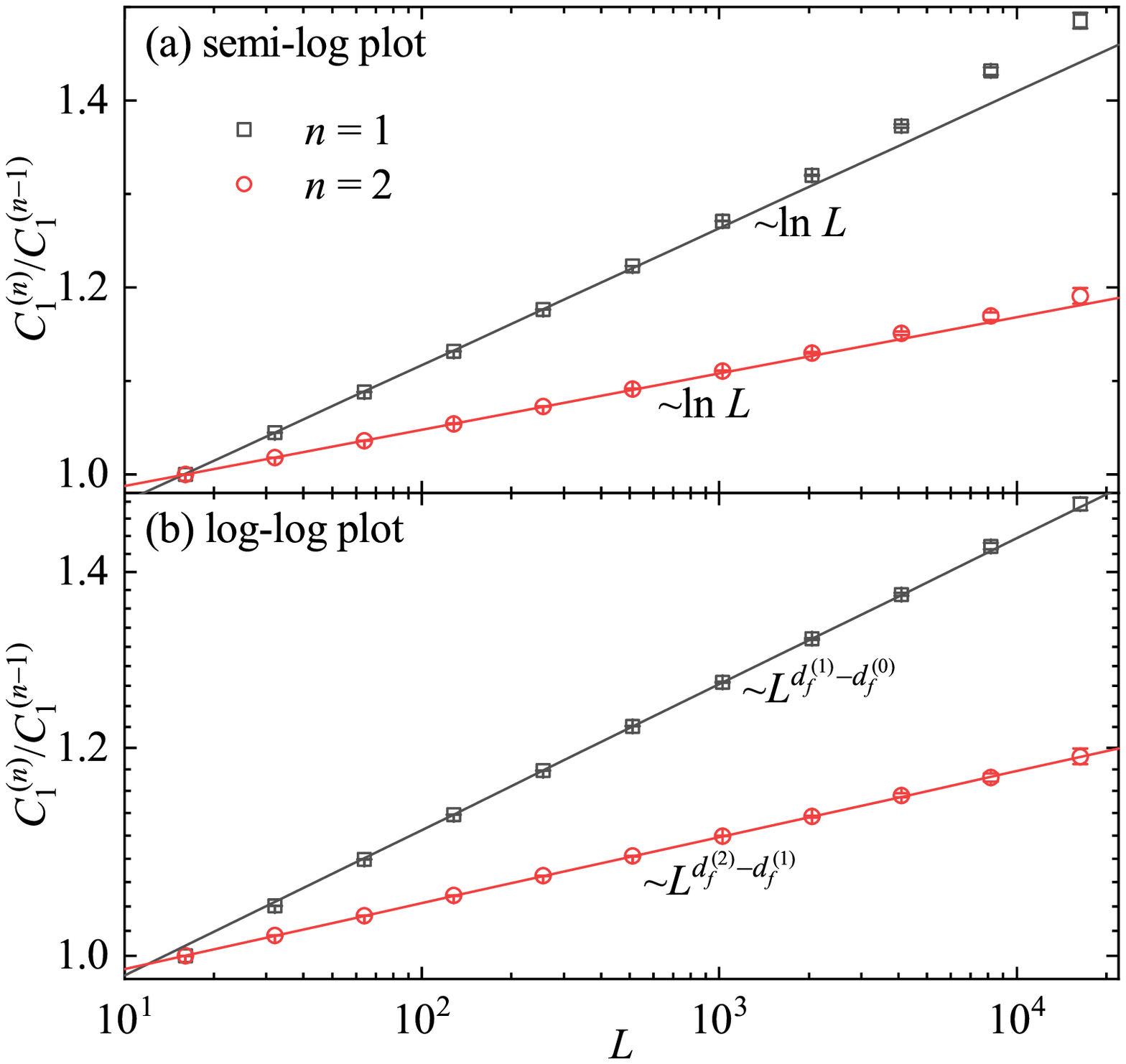}
\caption{(Color online) The finite-size scaling to demonstrate that the growth of $C_1^{(n)}/C_1^{(n-1)}$ for increasing system size is well described by a power-law function (b) rather than a logarithmic function (a), for generations $n=1$ and $2$. (a) In the semi-log plot, the ratio $C_1^{(n)}/C_1^{(n-1)}$ deviates upward from a straight line ($\sim\ln L$) for both $n=1$ and $2$, suggesting that the finite-size behaviors of $C_1^{(n)}/C_1^{(n-1)}$ cannot be described by logarithmic growth. (b) In the log-log plot, the same data as in (a) exhibit nice power-law behaviors indicated by the straight lines, representing the scaling $\sim L^{d_f^{\,(n)}-d_f^{\,(n-1)}}$ with the fractal dimensions from Table~\ref{t1}. To demonstrate generations $n=1$ and $2$ simultaneously, all the data are rescaled collectively to ensure that the first data point of different $n$ have the same value of $1$.} \label{f10}
\end{figure}

In a more comprehensive and complete discussion, it might be worthy to consider two additional potential outcomes when critical clusters merge. Firstly, a long-range order may be established, wherein a giant cluster immediately emerges, occupying a non-zero fraction of the entire lattice, leading the system out of criticality. Secondly, criticality may persist, with critical behaviors remaining unaffected, including unchanged critical exponents. The first scenario should yield $C_1\sim L^2$, while for the second scenario, $C_1$ consistently scales as $C_1\sim L^{d_f}$ with $d_f=91/48$. As clusters gradually grow larger over iterations, these two scaling forms might be influenced by a multiplicative logarithmic factor, and take forms of $C_1\sim L^2/\ln L$ and $C_1\sim L^{d_f} \ln L$ for the two scenarios, respectively. Due to this possibility, careful examinations of finite-size scaling should be conducted to avoid misidentifying a new fractal dimension from these multiplicative logarithmic corrections.

In the first scenario, the emergence of the giant cluster $C_1$ scaling as $\sim L^2$ after an iteration implies that the ratio $L^2/C_1^{(n)}$ should approach a constant as $L$ increases. When considering the multiplicative logarithmic correction $C_1\sim L^2/\ln L$, this ratio would exhibit a slow logarithmic growth with increasing $L$. However, as shown in Fig.~\ref{f9} (a), the finite-size scalings of $L^2/C_1^{(n)}$ for both $n=1$ and $2$ deviate upward from a logarithmic growth indicated by a straight line in the semi-log plot. This suggests that the growth of $L^2/C_1^{(n)}$ is faster than logarithmic. Furthermore, the same data collapse well onto a straight line in the log-log plot as shown in Fig.~\ref{f9} (b), representing the scaling behavior $L^2/C_1^{(n)}\sim L^{2-d_f^{(n)}}$. Thus, the generation-dependent fractal dimension is further supported, and the first scenario, i.e., a giant cluster emerges after an iteration, inadequately captures the physics of iterative percolation.

A similar analysis can be conducted for the second scenario by examining the finite-size behavior of the ratio $C_1^{(n)}/C_1^{(n-1)}$. If the second scenario holds true, indicating that all generations belong to the same universality class, the ratio $C_1^{(n)}/C_1^{(n-1)}$ would approach a constant or exhibit logarithmic growth. The data of $C_1^{(n)}/C_1^{(n-1)}$ for $n=1$ and $2$ with increasing $L$ are plotted in Fig.~\ref{f10} for both semi-log and log-log plots. Similar to Fig.~\ref{f9}, the finite-size scaling behavior of $C_1^{(n)}/C_1^{(n-1)}$ aligns entirely with power-law growth. Thus, this further confirms the generation-dependent fractal dimension and refutes the second scenario that the same universality holds for all generations.

\section{Bond-site percolation on iterative percolation clusters}  \label{app-bsp}

To further demonstrate the criticality of iterative site percolation, we introduce a dilution operation to drive the system deviating from criticality. Specifically, for any generation, we occupy the bonds between sites of the same color with probability $p_b$. We then consider the clusters connected by these occupied bonds as a bond-site percolation. For $p_b=1$, the bond-site percolation clusters correspond to the original configuration of iterative percolation. When $p_b<1$, the clusters from iterative percolation may break into smaller clusters. Therefore, by varying $p_b$, a percolation transition must occur for either white or black sites.

To determine the critical point $p_{b,c}$ of this bond-site percolation, we can study dimensionless quantities for either white or black clusters, such as the wrapping probability~\cite{Langlands1992,Pinson1994,Ziff1999}, the Binder cumulant~\cite{Binder1981a,Binder1981}, or the ratio of the two largest clusters, which take size-independent values at the critical point. Here, we employ the critical polynomial $R_P\equiv R_2-R_0$~\cite{Mertens2016,Xu2021}, where $R_2$ and $R_0$ denote the wrapping probabilities of having a cluster wrapping along both directions and having no wrapping cluster, respectively. For planar lattices, not only the critical values of $R_2$ and $R_0$ are universal, the relation $R_2=R_0$ always holds at criticality for the $2$D percolation universality. Therefore, the critical polynomial has the exact critical value $R_P(p_{b,c})=0$, for which finite-size corrections are minor or event absent.

\begin{figure}
\centering
\includegraphics[width=1.0\columnwidth]{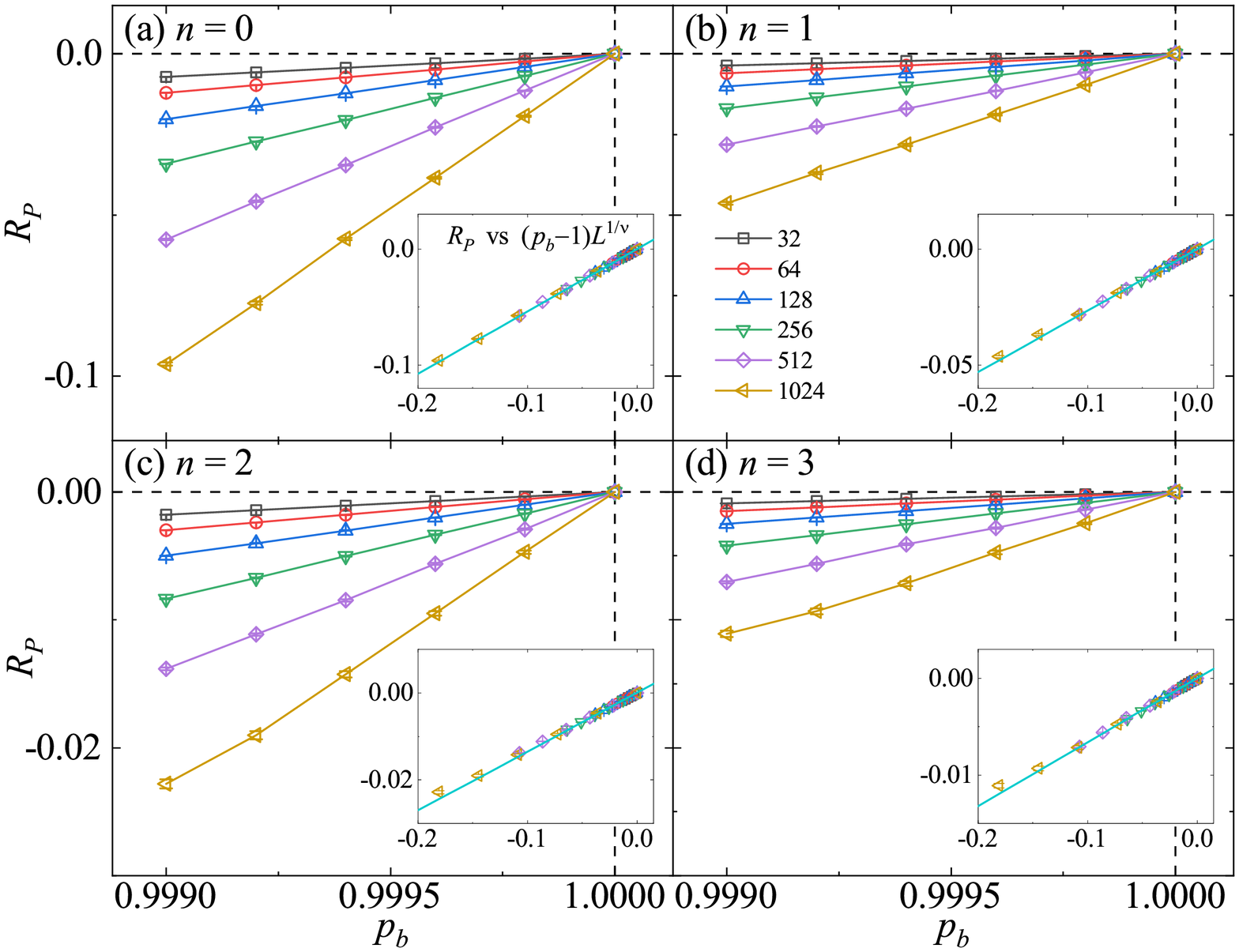}
\caption{(Color online) The critical polynomial $R_P$ for bond-site percolation on the iterative percolation configuration, for different generations $n$ and side lengths $L$, as a function of bond occupation probability $p_b$. In the simulation, the critical polynomial $R_P$ of a configuration is counted as the mean of those for white and black clusters. The crossing point of $R_P$ for different $L$ implies that this percolation model has a critical point at $p_{b,c}=1$, suggesting that the configuration of iterative percolation is at criticality. The insets plot $R_P$ as a function of $(p_b-1)L^{1/\nu}$ with $\nu=4/3$, where data of different system sizes collapse onto each other, indicated by the lines.} \label{f11}
\end{figure}

In Fig.~\ref{f11}, we plot $R_P(p_b, L)$ as a function of the probability $p_b$ for generations $n=0,1,2$, and $3$. We observe that for any generation, $R_P(p_b, L)$ for different $L$ intersect at the criticality $p_{b,c}=1$ with $R_{P,c}=0$. This further proves that the original configuration of iterative percolation is at criticality. In terms of renormalization group flow, the newly introduced parameter, bond occupied probability $p_b$, is a relevant parameter to the critical fixed point of iterative percolation; a small deviation of $p_b$ from $1$ can drive the system away from criticality.

Moreover, the self-matching property implies that in a white-black site configuration, when there is no wrapping cluster of one color along one direction, there must be a wrapping cluster of the opposite color along another direction. Therefore, for $p_b=1$, it must hold that $R_2^\text{black}=R_0^\text{white}$ and $R_2^\text{white}=R_0^\text{black}$, independent of system sizes. Thus, $R_P(1,L)$ in Fig.~\ref{f11}, which takes the average of white and black clusters in each white-black configuration, does not have any finite-size corrections, and error bars are absent for $p_b=1$.

\begin{table}
\caption{Fit results of bond-site percolation on the iterative percolation configuration. The fit results shown here are obtained by fitting the data of $R_P(p_b,L)$ to the scaling ansatz Eq.~(\ref{eq-pb}). Data without errors mean that they are fixed in the fit.}
\label{t2}
\begin{ruledtabular}
\begin{tabular}{cllll}
$n$  & \multicolumn{1}{c}{$1/\nu$}       & \multicolumn{1}{c}{$R_{P,c}$}     & \multicolumn{1}{c}{$p_{b,c}$}   & \multicolumn{1}{c}{$a_1$} \\
\hline
$0$  & $0.749(1)$    & $0.000001(8)$    & $1.000000(1)$    & $0.535(2)$  \\
     & $0.7494(7)$   & $0$              & $1$              & $0.535(1)$  \\
$1$  & $0.752(2)$    & $0.000004(5)$    & $0.999999(1)$    & $0.265(2)$  \\
     & $0.750(1)$    & $0$              & $1$              & $0.267(1)$  \\
$2$  & $0.743(3)$    & $0.000008(5)$    & $1.000008(6)$    & $0.137(1)$  \\
     & $0.747(2)$    & $0$              & $1$              & $0.135(1)$  \\
$3$  & $0.748(4)$    & $0.000000(2)$    & $1.000000(2)$    & $0.0672(9)$  \\
     & $0.748(2)$    & $0$              & $1$              & $0.0673(4)$
\end{tabular}
\end{ruledtabular}
\end{table}

We fit the data of $R_P(p_b,L)$ to the scaling ansatz
\begin{equation}
R_P(p_b,L)=R_{P,c} + a_1(p_b-p_{b,c})L^{1/\nu} + a_2(p_b-p_{b,c})^2L^{2/\nu},   \label{eq-pb}
\end{equation}
where $R_{P,c}$ denotes the critical value of $R_P$, and $\nu$ is the correlation-length exponent. With all terms free, the high-precision fit results in Table~\ref{t2} confirm that the critical point is generation-independent and always takes the value $p_{b,c}=1$, with $R_{P,c}=0$. Then, by fixing $R_{P,c}=0$ and $p_{b,c}=1$ in Eq.~(\ref{eq-pb}), the correlation-length exponent $\nu$ can be estimated with high precision. The results suggest a generation-independent correlation-length exponent $\nu$, taking the value $\nu=4/3$ for the $2$D percolation universality. By plotting $R_P$ as a function of $(p_b-p_{b,c})L^{1/\nu}$ with $p_{b,c}=1$ and $\nu=4/3$, the data of different system sizes collapse well onto each other (see the insets of Fig.~\ref{f11}). This reinforces that each generation has the critical point $p_{b,c}=1$ with the same correlation-length exponent $\nu=4/3$.

\bibliography{ref}

\begin{thebibliography}{43}%
\makeatletter
\providecommand \@ifxundefined [1]{%
 \@ifx{#1\undefined}
}%
\providecommand \@ifnum [1]{%
 \ifnum #1\expandafter \@firstoftwo
 \else \expandafter \@secondoftwo
 \fi
}%
\providecommand \@ifx [1]{%
 \ifx #1\expandafter \@firstoftwo
 \else \expandafter \@secondoftwo
 \fi
}%
\providecommand \natexlab [1]{#1}%
\providecommand \enquote  [1]{``#1''}%
\providecommand \bibnamefont  [1]{#1}%
\providecommand \bibfnamefont [1]{#1}%
\providecommand \citenamefont [1]{#1}%
\providecommand \href@noop [0]{\@secondoftwo}%
\providecommand \href [0]{\begingroup \@sanitize@url \@href}%
\providecommand \@href[1]{\@@startlink{#1}\@@href}%
\providecommand \@@href[1]{\endgroup#1\@@endlink}%
\providecommand \@sanitize@url [0]{\catcode `\\12\catcode `\$12\catcode
  `\&12\catcode `\#12\catcode `\^12\catcode `\_12\catcode `\%12\relax}%
\providecommand \@@startlink[1]{}%
\providecommand \@@endlink[0]{}%
\providecommand \url  [0]{\begingroup\@sanitize@url \@url }%
\providecommand \@url [1]{\endgroup\@href {#1}{\urlprefix }}%
\providecommand \urlprefix  [0]{URL }%
\providecommand \Eprint [0]{\href }%
\providecommand \doibase [0]{https://doi.org/}%
\providecommand \selectlanguage [0]{\@gobble}%
\providecommand \bibinfo  [0]{\@secondoftwo}%
\providecommand \bibfield  [0]{\@secondoftwo}%
\providecommand \translation [1]{[#1]}%
\providecommand \BibitemOpen [0]{}%
\providecommand \bibitemStop [0]{}%
\providecommand \bibitemNoStop [0]{.\EOS\space}%
\providecommand \EOS [0]{\spacefactor3000\relax}%
\providecommand \BibitemShut  [1]{\csname bibitem#1\endcsname}%
\let\auto@bib@innerbib\@empty
%</preamble>
\bibitem [{\citenamefont {Stauffer}\ and\ \citenamefont
  {Aharony}(1994)}]{Stauffer1991}%
  \BibitemOpen
  \bibfield  {author} {\bibinfo {author} {\bibfnamefont {D.}~\bibnamefont
  {Stauffer}}\ and\ \bibinfo {author} {\bibfnamefont {A.}~\bibnamefont
  {Aharony}},\ }\href@noop {} {\emph {\bibinfo {title} {Introduction to
  percolation theory}}},\ \bibinfo {edition} {revised 2nd}\ ed.\ (\bibinfo
  {publisher} {Taylor \& Francis},\ \bibinfo {address} {London},\ \bibinfo
  {year} {1994})\BibitemShut {NoStop}%
\bibitem [{\citenamefont {Baxter}(1989)}]{Baxter1989}%
  \BibitemOpen
  \bibfield  {author} {\bibinfo {author} {\bibfnamefont {R.~J.}\ \bibnamefont
  {Baxter}},\ }\href@noop {} {\emph {\bibinfo {title} {Exactly solved models in
  statistical mechanics}}}\ (\bibinfo  {publisher} {Academic Press},\ \bibinfo
  {address} {London},\ \bibinfo {year} {1989})\BibitemShut {NoStop}%
\bibitem [{\citenamefont {Sykes}\ and\ \citenamefont
  {Essam}(1964)}]{Sykes1964}%
  \BibitemOpen
  \bibfield  {author} {\bibinfo {author} {\bibfnamefont {M.~F.}\ \bibnamefont
  {Sykes}}\ and\ \bibinfo {author} {\bibfnamefont {J.~W.}\ \bibnamefont
  {Essam}},\ }\bibfield  {title} {\bibinfo {title} {Exact critical percolation
  probabilities for site and bond problems in two dimensions},\ }\href
  {https://doi.org/10.1063/1.1704215} {\bibfield  {journal} {\bibinfo
  {journal} {J. Math. Phys.}\ }\textbf {\bibinfo {volume} {5}},\ \bibinfo
  {pages} {1117} (\bibinfo {year} {1964})}\BibitemShut {NoStop}%
\bibitem [{\citenamefont {Ziff}\ and\ \citenamefont
  {Scullard}(2006)}]{Ziff2006}%
  \BibitemOpen
  \bibfield  {author} {\bibinfo {author} {\bibfnamefont {R.~M.}\ \bibnamefont
  {Ziff}}\ and\ \bibinfo {author} {\bibfnamefont {C.~R.}\ \bibnamefont
  {Scullard}},\ }\bibfield  {title} {\bibinfo {title} {Exact bond percolation
  thresholds in two dimensions},\ }\href
  {https://doi.org/10.1088/0305-4470/39/49/003} {\bibfield  {journal} {\bibinfo
   {journal} {J. Phys. A: Math. Gen.}\ }\textbf {\bibinfo {volume} {39}},\
  \bibinfo {pages} {15083} (\bibinfo {year} {2006})}\BibitemShut {NoStop}%
\bibitem [{\citenamefont {Lieb}(1967)}]{Lieb1967}%
  \BibitemOpen
  \bibfield  {author} {\bibinfo {author} {\bibfnamefont {E.~H.}\ \bibnamefont
  {Lieb}},\ }\bibfield  {title} {\bibinfo {title} {Exact solution of the
  problem of the entropy of two-dimensional ice},\ }\href
  {https://doi.org/10.1103/PhysRevLett.18.692} {\bibfield  {journal} {\bibinfo
  {journal} {Phys. Rev. Lett.}\ }\textbf {\bibinfo {volume} {18}},\ \bibinfo
  {pages} {692} (\bibinfo {year} {1967})}\BibitemShut {NoStop}%
\bibitem [{\citenamefont {Baxter}(1972)}]{Baxter1972}%
  \BibitemOpen
  \bibfield  {author} {\bibinfo {author} {\bibfnamefont {R.~J.}\ \bibnamefont
  {Baxter}},\ }\bibfield  {title} {\bibinfo {title} {Partition function of the
  eight-vertex lattice model},\ }\href
  {https://doi.org/10.1016/0003-4916(72)90335-1} {\bibfield  {journal}
  {\bibinfo  {journal} {Ann. Phys.}\ }\textbf {\bibinfo {volume} {70}},\
  \bibinfo {pages} {193} (\bibinfo {year} {1972})}\BibitemShut {NoStop}%
\bibitem [{\citenamefont {Belavin}\ \emph {et~al.}(1984)\citenamefont
  {Belavin}, \citenamefont {Polyakov},\ and\ \citenamefont
  {Zamolodchikov}}]{Belavin1984}%
  \BibitemOpen
  \bibfield  {author} {\bibinfo {author} {\bibfnamefont {A.~A.}\ \bibnamefont
  {Belavin}}, \bibinfo {author} {\bibfnamefont {A.~M.}\ \bibnamefont
  {Polyakov}},\ and\ \bibinfo {author} {\bibfnamefont {A.~B.}\ \bibnamefont
  {Zamolodchikov}},\ }\bibfield  {title} {\bibinfo {title} {Infinite conformal
  symmetry in two-dimensional quantum field theory},\ }\href
  {https://doi.org/10.1016/0550-3213(84)90052-x} {\bibfield  {journal}
  {\bibinfo  {journal} {Nucl. Phys. B}\ }\textbf {\bibinfo {volume} {241}},\
  \bibinfo {pages} {333} (\bibinfo {year} {1984})}\BibitemShut {NoStop}%
\bibitem [{\citenamefont {Friedan}\ \emph {et~al.}(1984)\citenamefont
  {Friedan}, \citenamefont {Qiu},\ and\ \citenamefont {Shenker}}]{Friedan1984}%
  \BibitemOpen
  \bibfield  {author} {\bibinfo {author} {\bibfnamefont {D.}~\bibnamefont
  {Friedan}}, \bibinfo {author} {\bibfnamefont {Z.}~\bibnamefont {Qiu}},\ and\
  \bibinfo {author} {\bibfnamefont {S.}~\bibnamefont {Shenker}},\ }\bibfield
  {title} {\bibinfo {title} {Conformal invariance, unitarity, and critical
  exponents in two dimensions},\ }\href
  {https://doi.org/10.1103/PhysRevLett.52.1575} {\bibfield  {journal} {\bibinfo
   {journal} {Phys. Rev. Lett.}\ }\textbf {\bibinfo {volume} {52}},\ \bibinfo
  {pages} {1575} (\bibinfo {year} {1984})}\BibitemShut {NoStop}%
\bibitem [{\citenamefont {Nienhuis}(1987)}]{Nienhuis1987}%
  \BibitemOpen
  \bibfield  {author} {\bibinfo {author} {\bibfnamefont {B.}~\bibnamefont
  {Nienhuis}},\ }\href@noop {} {\bibinfo {title} {Coulomb gas formulation of
  2{D} phase transition, in ``{P}hase transition and critical phenomena”, ed.
  {C. Domb}, {J. L. Lebowitz}}} (\bibinfo {year} {1987}),\ \bibinfo {note}
  {vol. 11}\BibitemShut {NoStop}%
\bibitem [{\citenamefont {Saleur}\ and\ \citenamefont
  {Duplantier}(1987)}]{Saleur1987}%
  \BibitemOpen
  \bibfield  {author} {\bibinfo {author} {\bibfnamefont {H.}~\bibnamefont
  {Saleur}}\ and\ \bibinfo {author} {\bibfnamefont {B.}~\bibnamefont
  {Duplantier}},\ }\bibfield  {title} {\bibinfo {title} {Exact determination of
  the percolation hull exponent in two dimensions},\ }\href
  {https://doi.org/10.1103/physrevlett.58.2325} {\bibfield  {journal} {\bibinfo
   {journal} {Phys. Rev. Lett.}\ }\textbf {\bibinfo {volume} {58}},\ \bibinfo
  {pages} {2325} (\bibinfo {year} {1987})}\BibitemShut {NoStop}%
\bibitem [{\citenamefont {Cardy}(1987)}]{Cardy1987}%
  \BibitemOpen
  \bibfield  {author} {\bibinfo {author} {\bibfnamefont {J.~L.}\ \bibnamefont
  {Cardy}},\ }\href@noop {} {\bibinfo {title} {Conformal invariance, in
  ``{P}hase transition and critical phenomena”, ed. {C. Domb}, {J. L.
  Lebowitz}}} (\bibinfo {year} {1987}),\ \bibinfo {note} {vol. 11}\BibitemShut
  {NoStop}%
\bibitem [{\citenamefont {Francesco}\ \emph {et~al.}(2012)\citenamefont
  {Francesco}, \citenamefont {Mathieu},\ and\ \citenamefont
  {S{\'e}n{\'e}chal}}]{Francesco2012}%
  \BibitemOpen
  \bibfield  {author} {\bibinfo {author} {\bibfnamefont {P.}~\bibnamefont
  {Francesco}}, \bibinfo {author} {\bibfnamefont {P.}~\bibnamefont {Mathieu}},\
  and\ \bibinfo {author} {\bibfnamefont {D.}~\bibnamefont {S{\'e}n{\'e}chal}},\
  }\href@noop {} {\emph {\bibinfo {title} {Conformal field theory}}}\ (\bibinfo
   {publisher} {Springer Science \& Business Media},\ \bibinfo {year}
  {2012})\BibitemShut {NoStop}%
\bibitem [{\citenamefont {Kager}\ and\ \citenamefont
  {Nienhuis}(2004)}]{Kager2004}%
  \BibitemOpen
  \bibfield  {author} {\bibinfo {author} {\bibfnamefont {W.}~\bibnamefont
  {Kager}}\ and\ \bibinfo {author} {\bibfnamefont {B.}~\bibnamefont
  {Nienhuis}},\ }\bibfield  {title} {\bibinfo {title} {A guide to stochastic
  {L}{\"o}wner evolution and its applications},\ }\href
  {https://doi.org/10.1023/b:joss.0000028058.87266.be} {\bibfield  {journal}
  {\bibinfo  {journal} {J. Stat. Phys.}\ }\textbf {\bibinfo {volume} {115}},\
  \bibinfo {pages} {1149} (\bibinfo {year} {2004})}\BibitemShut {NoStop}%
\bibitem [{\citenamefont {Cardy}(2005)}]{Cardy2005}%
  \BibitemOpen
  \bibfield  {author} {\bibinfo {author} {\bibfnamefont {J.}~\bibnamefont
  {Cardy}},\ }\bibfield  {title} {\bibinfo {title} {{SLE} for theoretical
  physicists},\ }\href {https://doi.org/10.1016/j.aop.2005.04.001} {\bibfield
  {journal} {\bibinfo  {journal} {Ann. Phys.}\ }\textbf {\bibinfo {volume}
  {318}},\ \bibinfo {pages} {81} (\bibinfo {year} {2005})}\BibitemShut
  {NoStop}%
\bibitem [{\citenamefont {Smirnov}\ and\ \citenamefont
  {Werner}(2001)}]{Smirnov2001}%
  \BibitemOpen
  \bibfield  {author} {\bibinfo {author} {\bibfnamefont {S.}~\bibnamefont
  {Smirnov}}\ and\ \bibinfo {author} {\bibfnamefont {W.}~\bibnamefont
  {Werner}},\ }\bibfield  {title} {\bibinfo {title} {Critical exponents for
  two-dimensional percolation},\ }\href
  {https://doi.org/10.4310/mrl.2001.v8.n6.a4} {\bibfield  {journal} {\bibinfo
  {journal} {Math. Res. Lett.}\ }\textbf {\bibinfo {volume} {8}},\ \bibinfo
  {pages} {729} (\bibinfo {year} {2001})}\BibitemShut {NoStop}%
\bibitem [{\citenamefont {Lawler}(2008)}]{Lawler2008}%
  \BibitemOpen
  \bibfield  {author} {\bibinfo {author} {\bibfnamefont {G.~F.}\ \bibnamefont
  {Lawler}},\ }\href@noop {} {\emph {\bibinfo {title} {Conformally invariant
  processes in the plane}}},\ \bibinfo {number} {114}\ (\bibinfo  {publisher}
  {American Mathematical Soc.},\ \bibinfo {year} {2008})\BibitemShut {NoStop}%
\bibitem [{\citenamefont {Jacobsen}(2009)}]{Jacobsen2009}%
  \BibitemOpen
  \bibfield  {author} {\bibinfo {author} {\bibfnamefont {J.~L.}\ \bibnamefont
  {Jacobsen}},\ }\bibfield  {title} {\bibinfo {title} {Conformal field theory
  applied to loop models},\ }in\ \href
  {https://doi.org/10.1007/978-1-4020-9927-4_14} {\emph {\bibinfo {booktitle}
  {Polygons, Polyominoes and Polycubes}}}\ (\bibinfo  {publisher} {Springer
  Netherlands},\ \bibinfo {year} {2009})\ pp.\ \bibinfo {pages}
  {347--424}\BibitemShut {NoStop}%
\bibitem [{\citenamefont {Newman}(2002)}]{Newman2002}%
  \BibitemOpen
  \bibfield  {author} {\bibinfo {author} {\bibfnamefont {M.~E.~J.}\
  \bibnamefont {Newman}},\ }\bibfield  {title} {\bibinfo {title} {Spread of
  epidemic disease on networks},\ }\href
  {https://doi.org/10.1103/physreve.66.016128} {\bibfield  {journal} {\bibinfo
  {journal} {Phys. Rev. E}\ }\textbf {\bibinfo {volume} {66}},\ \bibinfo
  {pages} {016128} (\bibinfo {year} {2002})}\BibitemShut {NoStop}%
\bibitem [{\citenamefont {Li}\ \emph {et~al.}(2014)\citenamefont {Li},
  \citenamefont {Fu}, \citenamefont {Wang}, \citenamefont {Lu}, \citenamefont
  {Berezin}, \citenamefont {Stanley},\ and\ \citenamefont {Havlin}}]{Li2014}%
  \BibitemOpen
  \bibfield  {author} {\bibinfo {author} {\bibfnamefont {D.}~\bibnamefont
  {Li}}, \bibinfo {author} {\bibfnamefont {B.}~\bibnamefont {Fu}}, \bibinfo
  {author} {\bibfnamefont {Y.}~\bibnamefont {Wang}}, \bibinfo {author}
  {\bibfnamefont {G.}~\bibnamefont {Lu}}, \bibinfo {author} {\bibfnamefont
  {Y.}~\bibnamefont {Berezin}}, \bibinfo {author} {\bibfnamefont {H.~E.}\
  \bibnamefont {Stanley}},\ and\ \bibinfo {author} {\bibfnamefont
  {S.}~\bibnamefont {Havlin}},\ }\bibfield  {title} {\bibinfo {title}
  {Percolation transition in dynamical traffic network with evolving critical
  bottlenecks},\ }\href {https://doi.org/10.1073/pnas.1419185112} {\bibfield
  {journal} {\bibinfo  {journal} {Proc. Natl. Acad. Sci.}\ }\textbf {\bibinfo
  {volume} {112}},\ \bibinfo {pages} {669} (\bibinfo {year}
  {2014})}\BibitemShut {NoStop}%
\bibitem [{\citenamefont {Fan}\ \emph {et~al.}(2018)\citenamefont {Fan},
  \citenamefont {Meng}, \citenamefont {Ashkenazy}, \citenamefont {Havlin},\
  and\ \citenamefont {Schellnhuber}}]{Fan2018}%
  \BibitemOpen
  \bibfield  {author} {\bibinfo {author} {\bibfnamefont {J.}~\bibnamefont
  {Fan}}, \bibinfo {author} {\bibfnamefont {J.}~\bibnamefont {Meng}}, \bibinfo
  {author} {\bibfnamefont {Y.}~\bibnamefont {Ashkenazy}}, \bibinfo {author}
  {\bibfnamefont {S.}~\bibnamefont {Havlin}},\ and\ \bibinfo {author}
  {\bibfnamefont {H.~J.}\ \bibnamefont {Schellnhuber}},\ }\bibfield  {title}
  {\bibinfo {title} {Climate network percolation reveals the expansion and
  weakening of the tropical component under global warming},\ }\href
  {https://doi.org/10.1073/pnas.1811068115} {\bibfield  {journal} {\bibinfo
  {journal} {Proc. Natl. Acad. Sci.}\ }\textbf {\bibinfo {volume} {115}},\
  \bibinfo {pages} {E12128} (\bibinfo {year} {2018})}\BibitemShut {NoStop}%
\bibitem [{\citenamefont {Xie}\ \emph {et~al.}(2021)\citenamefont {Xie},
  \citenamefont {Meng}, \citenamefont {Sun}, \citenamefont {Ma}, \citenamefont
  {Yan},\ and\ \citenamefont {Hu}}]{Xie2021}%
  \BibitemOpen
  \bibfield  {author} {\bibinfo {author} {\bibfnamefont {J.}~\bibnamefont
  {Xie}}, \bibinfo {author} {\bibfnamefont {F.}~\bibnamefont {Meng}}, \bibinfo
  {author} {\bibfnamefont {J.}~\bibnamefont {Sun}}, \bibinfo {author}
  {\bibfnamefont {X.}~\bibnamefont {Ma}}, \bibinfo {author} {\bibfnamefont
  {G.}~\bibnamefont {Yan}},\ and\ \bibinfo {author} {\bibfnamefont
  {Y.}~\bibnamefont {Hu}},\ }\bibfield  {title} {\bibinfo {title} {Detecting
  and modelling real percolation and phase transitions of information on social
  media},\ }\href {https://doi.org/10.1038/s41562-021-01090-z} {\bibfield
  {journal} {\bibinfo  {journal} {Nat. Hum. Behav.}\ }\textbf {\bibinfo
  {volume} {5}},\ \bibinfo {pages} {1161} (\bibinfo {year} {2021})}\BibitemShut
  {NoStop}%
\bibitem [{\citenamefont {Karrer}\ and\ \citenamefont
  {Newman}(2011)}]{Karrer2011}%
  \BibitemOpen
  \bibfield  {author} {\bibinfo {author} {\bibfnamefont {B.}~\bibnamefont
  {Karrer}}\ and\ \bibinfo {author} {\bibfnamefont {M.~E.~J.}\ \bibnamefont
  {Newman}},\ }\bibfield  {title} {\bibinfo {title} {Competing epidemics on
  complex networks},\ }\href {https://doi.org/10.1103/physreve.84.036106}
  {\bibfield  {journal} {\bibinfo  {journal} {Phys. Rev. E}\ }\textbf {\bibinfo
  {volume} {84}},\ \bibinfo {pages} {036106} (\bibinfo {year}
  {2011})}\BibitemShut {NoStop}%
\bibitem [{\citenamefont {Hu}\ \emph {et~al.}(2011)\citenamefont {Hu},
  \citenamefont {Deng},\ and\ \citenamefont {Bl\"ote}}]{Hu2011}%
  \BibitemOpen
  \bibfield  {author} {\bibinfo {author} {\bibfnamefont {H.}~\bibnamefont
  {Hu}}, \bibinfo {author} {\bibfnamefont {Y.}~\bibnamefont {Deng}},\ and\
  \bibinfo {author} {\bibfnamefont {H.~W.~J.}\ \bibnamefont {Bl\"ote}},\
  }\bibfield  {title} {\bibinfo {title} {Berezinskii-kosterlitz-thouless-like
  percolation transitions in the two-dimensional $\mathit{XY}$ model},\ }\href
  {https://doi.org/10.1103/PhysRevE.83.011124} {\bibfield  {journal} {\bibinfo
  {journal} {Phys. Rev. E}\ }\textbf {\bibinfo {volume} {83}},\ \bibinfo
  {pages} {011124} (\bibinfo {year} {2011})}\BibitemShut {NoStop}%
\bibitem [{\citenamefont {Granell}\ \emph {et~al.}(2014)\citenamefont
  {Granell}, \citenamefont {G{\'{o}}mez},\ and\ \citenamefont
  {Arenas}}]{Granell2014}%
  \BibitemOpen
  \bibfield  {author} {\bibinfo {author} {\bibfnamefont {C.}~\bibnamefont
  {Granell}}, \bibinfo {author} {\bibfnamefont {S.}~\bibnamefont
  {G{\'{o}}mez}},\ and\ \bibinfo {author} {\bibfnamefont {A.}~\bibnamefont
  {Arenas}},\ }\bibfield  {title} {\bibinfo {title} {Competing spreading
  processes on multiplex networks: {A}wareness and epidemics},\ }\href
  {https://doi.org/10.1103/physreve.90.012808} {\bibfield  {journal} {\bibinfo
  {journal} {Phys. Rev. E}\ }\textbf {\bibinfo {volume} {90}},\ \bibinfo
  {pages} {012808} (\bibinfo {year} {2014})}\BibitemShut {NoStop}%
\bibitem [{\citenamefont {Massaro}\ and\ \citenamefont
  {Bagnoli}(2014)}]{Massaro2014}%
  \BibitemOpen
  \bibfield  {author} {\bibinfo {author} {\bibfnamefont {E.}~\bibnamefont
  {Massaro}}\ and\ \bibinfo {author} {\bibfnamefont {F.}~\bibnamefont
  {Bagnoli}},\ }\bibfield  {title} {\bibinfo {title} {Epidemic spreading and
  risk perception in multiplex networks: {A} self-organized percolation
  method},\ }\href {https://doi.org/10.1103/physreve.90.052817} {\bibfield
  {journal} {\bibinfo  {journal} {Phys. Rev. E}\ }\textbf {\bibinfo {volume}
  {90}},\ \bibinfo {pages} {052817} (\bibinfo {year} {2014})}\BibitemShut
  {NoStop}%
\bibitem [{\citenamefont {Liu}\ \emph {et~al.}(2015)\citenamefont {Liu},
  \citenamefont {Deng},\ and\ \citenamefont {Jacobsen}}]{Liu2015}%
  \BibitemOpen
  \bibfield  {author} {\bibinfo {author} {\bibfnamefont {X.-W.}\ \bibnamefont
  {Liu}}, \bibinfo {author} {\bibfnamefont {Y.}~\bibnamefont {Deng}},\ and\
  \bibinfo {author} {\bibfnamefont {J.~L.}\ \bibnamefont {Jacobsen}},\
  }\bibfield  {title} {\bibinfo {title} {Recursive percolation},\ }\href
  {https://doi.org/10.1103/PhysRevE.92.010103} {\bibfield  {journal} {\bibinfo
  {journal} {Phys. Rev. E}\ }\textbf {\bibinfo {volume} {92}},\ \bibinfo
  {pages} {010103(R)} (\bibinfo {year} {2015})}\BibitemShut {NoStop}%
\bibitem [{\citenamefont {Hu}\ \emph {et~al.}(2016)\citenamefont {Hu},
  \citenamefont {Ziff},\ and\ \citenamefont {Deng}}]{Hu2016}%
  \BibitemOpen
  \bibfield  {author} {\bibinfo {author} {\bibfnamefont {H.}~\bibnamefont
  {Hu}}, \bibinfo {author} {\bibfnamefont {R.~M.}\ \bibnamefont {Ziff}},\ and\
  \bibinfo {author} {\bibfnamefont {Y.}~\bibnamefont {Deng}},\ }\bibfield
  {title} {\bibinfo {title} {No-enclave percolation corresponds to holes in the
  cluster backbone},\ }\href {https://doi.org/10.1103/PhysRevLett.117.185701}
  {\bibfield  {journal} {\bibinfo  {journal} {Phys. Rev. Lett.}\ }\textbf
  {\bibinfo {volume} {117}},\ \bibinfo {pages} {185701} (\bibinfo {year}
  {2016})}\BibitemShut {NoStop}%
\bibitem [{\citenamefont {Li}\ \emph {et~al.}(2020)\citenamefont {Li},
  \citenamefont {L\"{u}}, \citenamefont {Deng}, \citenamefont {Hu},
  \citenamefont {Wang}, \citenamefont {Medo},\ and\ \citenamefont
  {Stanley}}]{Li2020}%
  \BibitemOpen
  \bibfield  {author} {\bibinfo {author} {\bibfnamefont {M.}~\bibnamefont
  {Li}}, \bibinfo {author} {\bibfnamefont {L.}~\bibnamefont {L\"{u}}}, \bibinfo
  {author} {\bibfnamefont {Y.}~\bibnamefont {Deng}}, \bibinfo {author}
  {\bibfnamefont {M.-B.}\ \bibnamefont {Hu}}, \bibinfo {author} {\bibfnamefont
  {H.}~\bibnamefont {Wang}}, \bibinfo {author} {\bibfnamefont {M.}~\bibnamefont
  {Medo}},\ and\ \bibinfo {author} {\bibfnamefont {H.~E.}\ \bibnamefont
  {Stanley}},\ }\bibfield  {title} {\bibinfo {title} {History-dependent
  percolation on multiplex networks},\ }\href
  {https://doi.org/10.1093/nsr/nwaa029} {\bibfield  {journal} {\bibinfo
  {journal} {Natl. Sci. Rev.}\ }\textbf {\bibinfo {volume} {7}},\ \bibinfo
  {pages} {1296} (\bibinfo {year} {2020})}\BibitemShut {NoStop}%
\bibitem [{\citenamefont {Hu}\ \emph {et~al.}(2020)\citenamefont {Hu},
  \citenamefont {Sun}, \citenamefont {Wang}, \citenamefont {Lv},\ and\
  \citenamefont {Deng}}]{Hu2020}%
  \BibitemOpen
  \bibfield  {author} {\bibinfo {author} {\bibfnamefont {M.}~\bibnamefont
  {Hu}}, \bibinfo {author} {\bibfnamefont {Y.}~\bibnamefont {Sun}}, \bibinfo
  {author} {\bibfnamefont {D.}~\bibnamefont {Wang}}, \bibinfo {author}
  {\bibfnamefont {J.-P.}\ \bibnamefont {Lv}},\ and\ \bibinfo {author}
  {\bibfnamefont {Y.}~\bibnamefont {Deng}},\ }\bibfield  {title} {\bibinfo
  {title} {History-dependent percolation in two dimensions},\ }\href
  {https://doi.org/10.1103/PhysRevE.102.052121} {\bibfield  {journal} {\bibinfo
   {journal} {Phys. Rev. E}\ }\textbf {\bibinfo {volume} {102}},\ \bibinfo
  {pages} {052121} (\bibinfo {year} {2020})}\BibitemShut {NoStop}%
\bibitem [{\citenamefont {Havlin}\ \emph {et~al.}(1983)\citenamefont {Havlin},
  \citenamefont {Ben-Avraham},\ and\ \citenamefont {Movshovitz}}]{Havlin1983}%
  \BibitemOpen
  \bibfield  {author} {\bibinfo {author} {\bibfnamefont {S.}~\bibnamefont
  {Havlin}}, \bibinfo {author} {\bibfnamefont {D.}~\bibnamefont
  {Ben-Avraham}},\ and\ \bibinfo {author} {\bibfnamefont {D.}~\bibnamefont
  {Movshovitz}},\ }\bibfield  {title} {\bibinfo {title} {Percolation on fractal
  lattices},\ }\href {https://doi.org/10.1103/physrevlett.51.2347} {\bibfield
  {journal} {\bibinfo  {journal} {Phys. Rev. Lett.}\ }\textbf {\bibinfo
  {volume} {51}},\ \bibinfo {pages} {2347} (\bibinfo {year}
  {1983})}\BibitemShut {NoStop}%
\bibitem [{\citenamefont {Ma}(2018)}]{Ma2018}%
  \BibitemOpen
  \bibfield  {author} {\bibinfo {author} {\bibfnamefont {S.-K.}\ \bibnamefont
  {Ma}},\ }\href@noop {} {\emph {\bibinfo {title} {Modern theory of critical
  phenomena}}}\ (\bibinfo  {publisher} {Routledge},\ \bibinfo {address} {New
  York},\ \bibinfo {year} {2018})\BibitemShut {NoStop}%
\bibitem [{\citenamefont {Duplantier}(1989)}]{Duplantier1989}%
  \BibitemOpen
  \bibfield  {author} {\bibinfo {author} {\bibfnamefont {B.}~\bibnamefont
  {Duplantier}},\ }\bibfield  {title} {\bibinfo {title} {Two-dimensional
  fractal geometry, critical phenomena and conformal invariance},\ }\href
  {https://doi.org/10.1016/0370-1573(89)90042-2} {\bibfield  {journal}
  {\bibinfo  {journal} {Phys. Rep.}\ }\textbf {\bibinfo {volume} {184}},\
  \bibinfo {pages} {229} (\bibinfo {year} {1989})}\BibitemShut {NoStop}%
\bibitem [{\citenamefont {Wittmann}\ and\ \citenamefont
  {Young}(2014)}]{Wittmann2014}%
  \BibitemOpen
  \bibfield  {author} {\bibinfo {author} {\bibfnamefont {M.}~\bibnamefont
  {Wittmann}}\ and\ \bibinfo {author} {\bibfnamefont {A.~P.}\ \bibnamefont
  {Young}},\ }\bibfield  {title} {\bibinfo {title} {Finite-size scaling above
  the upper critical dimension},\ }\href
  {https://doi.org/10.1103/physreve.90.062137} {\bibfield  {journal} {\bibinfo
  {journal} {Phys. Rev. E}\ }\textbf {\bibinfo {volume} {90}},\ \bibinfo
  {pages} {062137} (\bibinfo {year} {2014})}\BibitemShut {NoStop}%
\bibitem [{\citenamefont {Maes}\ and\ \citenamefont {Velde}(1995)}]{Maes1995}%
  \BibitemOpen
  \bibfield  {author} {\bibinfo {author} {\bibfnamefont {C.}~\bibnamefont
  {Maes}}\ and\ \bibinfo {author} {\bibfnamefont {K.~V.}\ \bibnamefont
  {Velde}},\ }\bibfield  {title} {\bibinfo {title} {The fuzzy {Potts} model},\
  }\href {https://doi.org/10.1088/0305-4470/28/15/007} {\bibfield  {journal}
  {\bibinfo  {journal} {J. Phys. A: Math. Gen.}\ }\textbf {\bibinfo {volume}
  {28}},\ \bibinfo {pages} {4261} (\bibinfo {year} {1995})}\BibitemShut
  {NoStop}%
\bibitem [{\citenamefont {H\"{a}ggstr\"{o}m}(2001)}]{Haeggstroem2001}%
  \BibitemOpen
  \bibfield  {author} {\bibinfo {author} {\bibfnamefont {O.}~\bibnamefont
  {H\"{a}ggstr\"{o}m}},\ }\bibfield  {title} {\bibinfo {title} {Coloring
  percolation clusters at random},\ }\href
  {https://doi.org/10.1016/s0304-4149(01)00115-6} {\bibfield  {journal}
  {\bibinfo  {journal} {Stoch. Proc. Appl.}\ }\textbf {\bibinfo {volume}
  {96}},\ \bibinfo {pages} {213} (\bibinfo {year} {2001})}\BibitemShut
  {NoStop}%
\bibitem [{\citenamefont {Nolin}\ \emph {et~al.}(2023)\citenamefont {Nolin},
  \citenamefont {Qian}, \citenamefont {Sun},\ and\ \citenamefont
  {Zhuang}}]{Nolin2023}%
  \BibitemOpen
  \bibfield  {author} {\bibinfo {author} {\bibfnamefont {P.}~\bibnamefont
  {Nolin}}, \bibinfo {author} {\bibfnamefont {W.}~\bibnamefont {Qian}},
  \bibinfo {author} {\bibfnamefont {X.}~\bibnamefont {Sun}},\ and\ \bibinfo
  {author} {\bibfnamefont {Z.}~\bibnamefont {Zhuang}},\ }\bibfield  {title}
  {\bibinfo {title} {Backbone exponent for two-dimensional percolation},\
  }\href {https://arxiv.org/abs/2309.05050} {\bibfield  {journal} {\bibinfo
  {journal} {arXiv preprint}\ ,\ \bibinfo {pages} {arXiv:2309.05050}} (\bibinfo
  {year} {2023})}\BibitemShut {NoStop}%
\bibitem [{\citenamefont {Langlands}\ \emph {et~al.}(1992)\citenamefont
  {Langlands}, \citenamefont {Pichet}, \citenamefont {Pouliot},\ and\
  \citenamefont {Saint-Aubin}}]{Langlands1992}%
  \BibitemOpen
  \bibfield  {author} {\bibinfo {author} {\bibfnamefont {R.~P.}\ \bibnamefont
  {Langlands}}, \bibinfo {author} {\bibfnamefont {C.}~\bibnamefont {Pichet}},
  \bibinfo {author} {\bibfnamefont {P.}~\bibnamefont {Pouliot}},\ and\ \bibinfo
  {author} {\bibfnamefont {Y.}~\bibnamefont {Saint-Aubin}},\ }\bibfield
  {title} {\bibinfo {title} {On the universality of crossing probabilities in
  two-dimensional percolation},\ }\href {https://doi.org/10.1007/bf01049720}
  {\bibfield  {journal} {\bibinfo  {journal} {J. Stat. Phys.}\ }\textbf
  {\bibinfo {volume} {67}},\ \bibinfo {pages} {553} (\bibinfo {year}
  {1992})}\BibitemShut {NoStop}%
\bibitem [{\citenamefont {Pinson}(1994)}]{Pinson1994}%
  \BibitemOpen
  \bibfield  {author} {\bibinfo {author} {\bibfnamefont {H.~T.}\ \bibnamefont
  {Pinson}},\ }\bibfield  {title} {\bibinfo {title} {Critical percolation on
  the torus},\ }\href {https://doi.org/10.1007/bf02186762} {\bibfield
  {journal} {\bibinfo  {journal} {J. Stat. Phys.}\ }\textbf {\bibinfo {volume}
  {75}},\ \bibinfo {pages} {1167} (\bibinfo {year} {1994})}\BibitemShut
  {NoStop}%
\bibitem [{\citenamefont {Ziff}\ \emph {et~al.}(1999)\citenamefont {Ziff},
  \citenamefont {Lorenz},\ and\ \citenamefont {Kleban}}]{Ziff1999}%
  \BibitemOpen
  \bibfield  {author} {\bibinfo {author} {\bibfnamefont {R.~M.}\ \bibnamefont
  {Ziff}}, \bibinfo {author} {\bibfnamefont {C.~D.}\ \bibnamefont {Lorenz}},\
  and\ \bibinfo {author} {\bibfnamefont {P.}~\bibnamefont {Kleban}},\
  }\bibfield  {title} {\bibinfo {title} {Shape-dependent universality in
  percolation},\ }\href {https://doi.org/10.1016/s0378-4371(98)00569-x}
  {\bibfield  {journal} {\bibinfo  {journal} {Physica A}\ }\textbf {\bibinfo
  {volume} {266}},\ \bibinfo {pages} {17} (\bibinfo {year} {1999})}\BibitemShut
  {NoStop}%
\bibitem [{\citenamefont {Binder}(1981{\natexlab{a}})}]{Binder1981a}%
  \BibitemOpen
  \bibfield  {author} {\bibinfo {author} {\bibfnamefont {K.}~\bibnamefont
  {Binder}},\ }\bibfield  {title} {\bibinfo {title} {Critical properties from
  monte carlo coarse graining and renormalization},\ }\href
  {https://doi.org/10.1103/physrevlett.47.693} {\bibfield  {journal} {\bibinfo
  {journal} {Phys. Rev. Lett.}\ }\textbf {\bibinfo {volume} {47}},\ \bibinfo
  {pages} {693} (\bibinfo {year} {1981}{\natexlab{a}})}\BibitemShut {NoStop}%
\bibitem [{\citenamefont {Binder}(1981{\natexlab{b}})}]{Binder1981}%
  \BibitemOpen
  \bibfield  {author} {\bibinfo {author} {\bibfnamefont {K.}~\bibnamefont
  {Binder}},\ }\bibfield  {title} {\bibinfo {title} {Finite size scaling
  analysis of {Ising} model block distribution functions},\ }\href
  {https://doi.org/10.1007/bf01293604} {\bibfield  {journal} {\bibinfo
  {journal} {Z. Physik B - Condensed Matter}\ }\textbf {\bibinfo {volume}
  {43}},\ \bibinfo {pages} {119} (\bibinfo {year}
  {1981}{\natexlab{b}})}\BibitemShut {NoStop}%
\bibitem [{\citenamefont {Mertens}\ and\ \citenamefont
  {Ziff}(2016)}]{Mertens2016}%
  \BibitemOpen
  \bibfield  {author} {\bibinfo {author} {\bibfnamefont {S.}~\bibnamefont
  {Mertens}}\ and\ \bibinfo {author} {\bibfnamefont {R.~M.}\ \bibnamefont
  {Ziff}},\ }\bibfield  {title} {\bibinfo {title} {Percolation in finite
  matching lattices},\ }\href {https://doi.org/10.1103/physreve.94.062152}
  {\bibfield  {journal} {\bibinfo  {journal} {Phys. Rev. E}\ }\textbf {\bibinfo
  {volume} {94}},\ \bibinfo {pages} {062152} (\bibinfo {year}
  {2016})}\BibitemShut {NoStop}%
\bibitem [{\citenamefont {Xu}\ \emph {et~al.}(2021)\citenamefont {Xu},
  \citenamefont {Wang}, \citenamefont {Hu},\ and\ \citenamefont
  {Deng}}]{Xu2021}%
  \BibitemOpen
  \bibfield  {author} {\bibinfo {author} {\bibfnamefont {W.}~\bibnamefont
  {Xu}}, \bibinfo {author} {\bibfnamefont {J.}~\bibnamefont {Wang}}, \bibinfo
  {author} {\bibfnamefont {H.}~\bibnamefont {Hu}},\ and\ \bibinfo {author}
  {\bibfnamefont {Y.}~\bibnamefont {Deng}},\ }\bibfield  {title} {\bibinfo
  {title} {Critical polynomials in the nonplanar and continuum percolation
  models},\ }\href {https://doi.org/10.1103/physreve.103.022127} {\bibfield
  {journal} {\bibinfo  {journal} {Phys. Rev. E}\ }\textbf {\bibinfo {volume}
  {103}},\ \bibinfo {pages} {022127} (\bibinfo {year} {2021})}\BibitemShut
  {NoStop}%
\end{thebibliography}%

\end{document}